\documentclass[pra,aps,superbib,citeautoscript,twocolumn]{revtex4-1}

\usepackage[colorlinks=true,urlcolor=blue,citecolor=blue,linkcolor=blue]{hyperref}

\usepackage{amsmath,bm}
\usepackage{amstext}
\usepackage{epsfig}
\usepackage{xcolor}
\usepackage{subfig}
\usepackage{graphicx,amsmath,amssymb,tabularx}
\usepackage{multirow}
\usepackage{array}
\usepackage{dsfont}
\usepackage{caption}
\usepackage{cleveref}
\usepackage{ulem}
\usepackage[toc]{appendix}
\usepackage{physics}
\usepackage{mathrsfs}  
\usepackage{amsbsy}

\definecolor{dgreen}{rgb}{0,.5,0}
\definecolor{dred}{rgb}{.7,.0,.0}

\DeclareMathAlphabet\mathbfcal{OMS}{cmsy}{b}{n}

\newcommand{\bxi}{\bm{\xi}}

\newcommand{\bze}{\bm{\zeta}}
\newcommand{\br}{\mathbf{r}}

\newcommand{\ie}{{\it i.e.}}
\newcommand{\myket}[1]{\left\vert #1\right\rangle}
\newcommand{\mybra}[1]{\left\langle #1\right\vert}
\newcommand{\inner}[2]{\left\langle #1 \middle\vert #2 \right\rangle}
\newcommand{\innerop}[3]{\left\langle #1 \middle\vert #2 \middle\vert #3 \right\rangle}
\newcommand{\contract}[2]{\left( #1 \middle\vert #2 \right)}

\newcommand{\manu}[1]{{\textcolor{blue}{Manu: #1}} }

\newcommand{\be}{\begin{eqnarray}}
\newcommand{\ee}{\end{eqnarray}}
\newcommand{\bse}{\begin{subequations}}
\newcommand{\ese}{\end{subequations}}
\DeclareMathAlphabet\mathbfcal{OMS}{cmsy}{b}{n}

\begin{document}

\title{
Ensemble density functional theory of ground and excited energy levels
}

\author{
Emmanuel Fromager
}
\email{fromagere@unistra.fr}
\affiliation{\it 
~\\
Laboratoire de Chimie Quantique,
Institut de Chimie, CNRS / Universit\'{e} de Strasbourg,
4 rue Blaise Pascal, 67000 Strasbourg, France\\
\\
}
\affiliation{\it
~\\
University of Strasbourg Institute for Advanced Study,  
5, all\'{e}e du G\'{e}n\'{e}ral Rouvillois, F-67083 Strasbourg, France
\\
}


\begin{abstract}
A Kohn--Sham density-functional energy expression is derived for any
(ground or excited) state within a given many-electron ensemble along with the 
stationarity condition it fulfills with respect to the ensemble density, thus giving access to
both physical energy levels and individual-state densities, in principle
exactly. 
We also provide working equations for the evaluation of the latter from
the true static ensemble
density-density linear response function. Unlike in Gould's recent
ensemble potential functional approach to excited states 
[arXiv:2404.12593], we use the ensemble density as sole basic variable.   
While a state-specific KS
potential naturally emerges from the present formalism, at the exact ensemble
Hartree-exchange-only (Hx) level of approximation, the standard
implementation of   
orbital-optimized DFT for excited states is recovered when recycling the regular
ground-state Hx-correlation functional in this context. 
\end{abstract}

\maketitle



\section{Introduction}

Density-functional theory (DFT)~\cite{hktheo,KS} offers a drastic simplification of the
ground-state electronic structure problem by mapping, in principle
exactly, the electronic density
onto a fictitious noninteracting (so-called Kohn--Sham (KS)) electronic system. The
true (interacting) energy can then be determined from the universal
Hartree-exchange-correlation (Hxc) density-functional energy
contribution for which ever more accurate approximations are routinely used. Such a setting allows for large-scale
computations, which explains why DFT has become the workhorse of quantum
chemistry and materials science~\cite{Teale2022_DFT_exchange}. Extending DFT to the excited
states is not trivial since, unlike the ground-state energy,
excited-state energies are not (local) minima of the electronic energy.
In this respect, the extension of DFT to the time-dependent (TD) regime
is a very appealing approach since the dynamical linear response of the
density gives access to the excitation energies, in principle
exactly~\cite{runge1984density,casida1995timedependent,Casida_tddft_review_2012,Lacombe2023_Non-adiabatic}.
Despite its success, linear response TDDFT still suffers from various
deficiencies. The absence of memory effects in regular Hxc functionals
prevents the description of multiple electronic excitations, for
example~\cite{maitra2004double,cave2004dressed,Huix-Rotllant2011_Assessment,elliott2011perspectives,maitra2022double,Casida_tddft_review_2012,Lacombe2023_Non-adiabatic}.
Moreover, the single-reference linear response setting is {\it a priori} not
adequate for dealing with (quasi-) degenerate situations, like in the
vicinity of a conical intersection~\cite{Casida_tddft_review_2012}. This is the reason why alternative
time-independent DFTs of excited states have continued to be developed
over the years at
both formal and practical
levels, either from an ensemble
perspective~\cite{Fan1949_On,JPC79_Theophilou_equi-ensembles,Hendekovic1982_equi-ensembles,gross1988rayleigh, gross1988density,oliveira1988density,deur2017exact,yang2017direct,gould2017hartree,gould2018charge,deur2018exploring,
PRL19_Gould_DD_correlation,Fromager_2020,PRL20_Gould_Hartree_def_from_ACDF_th,
Gould2020_Approximately,
loos2020weightdependent,Yang2021_Second,Gould2021_Ensemble_ugly,gould2023local,Gould2023_Electronic,gould2022single,
Gould2021_Double,
Cernatic2022,Schilling2021_Ensemble,Liebert2022_Foundation,Benavides-Riveros2022_Excitations,
Liebert_2023_An_exact_bosons,Liebert2023_Deriving,ding2024ground,Scott2024_Exact,Cernatic2024_Neutral,cernatic2024extended_doubles,gould2024excitationenergiesstatespecificensemble} or a state-specific
one~\cite{Gorling99_Density,ayers2009pra,Gilbert08_Self-Consistent_MOM,Besley09_Self-consistent_MUM,Barca18_Simple_MOM,JCP09_Ziegler_relation_TD-DFT_VDFT,JCTC13_Ziegler_SCF-CV-DFT,levy2016computation,PRA12_Nagy_TinD-DFT_ES,JCP15_Ayers_KS-DFT_excit-states_Coulomb,Ayers2018_Article_Time-independentDensityFunctio,Garrigue2022,Giarrusso2023_Exact}.\\

In recent years, the orbital-optimized DFT computation of excited states
has shown very promising
results~\cite{Levi20_Variational,Ivanov21_Method,Hait21_Orbital,Schmerwitz22_Variational}.
Unlike KS-DFT for ensembles~\cite{gross1988density}, which has
solid foundations like regular ground-state
KS-DFT~\cite{gross1988rayleigh,Teale2022_DFT_exchange}, such computational
strategies have no obvious from-first-principles justification, simply
because the Rayleigh--Ritz variational principle does not hold for
(individual) excited states. Very recently, Yang and Ayers~\cite{yang2024foundationdeltascfapproachdensity} derived a KS theory of ground and excited
states where, among other possible choices, a noninteracting
wavefunction can be used as basic variable, thus filling an important
gap between the exact theory and practical computational
implementations. Formulating a proper stationarity
condition is central in such an exactification process. In ensemble DFT, the ensemble energy is stationary with respect
to the ensemble density~\cite{gross1988density}. On the other hand, the
extraction of a given energy level
from the former, which is a post ensemble KS-DFT
treatment~\cite{deur2019ground}, is not a variational procedure~\cite{Fromager_2020}. Nevertheless,
Gould~\cite{gould2024stationaryconditionsexcitedstates} has recently
derived an ensemble {\it potential} functional theory where individual
energy levels can be identified as stationary points. This key 
result, which offers an alternative (and still exact, in principle)
approach to DFT-based calculations of excited states, raises several
fundamental and practical questions. For example, can we derive a similar theory that
uses the ensemble {\it density} as sole basic variable instead? This
would make the connection with regular (ground-state) DFT even clearer,
which is important for rationalizing and possibly improving the use of standard
(ground-state) DFT functionals in excited-state calculations. Will a
KS-like equation naturally emerge from the resulting ensemble density-functional stationarity
condition? Is it possible to identify, in such a setting, the ensemble density-functional
approximations that underly current DFT computations of excited
states?\\ 

The purpose of the present work is to address the above questions. The
paper is organized as follows. After a brief review of ensemble DFT for
neutral electronic excitations (in Sec.~\ref{sec:review_eDFT}), the
stationarity of ground and excited energy levels (within the ensemble) with respect to the
ensemble density is established (in
Sec.~\ref{sec:ens_dens_func_energy_levels_and_stat_cond}). In addition,
a KS decomposition is introduced for each ensemble density-functional energy level 
where individual Hxc
functionals are connected explicitly to the
ensemble one. The implications of each energy level stationarity, when
combined with this KS decomposition, are discussed in detail in
Sec.~\ref{sec:ind_state_KSDFT}, with a particular focus on how exact
individual-state densities emerge from the theory. Working equations (where the true static ensemble density-density linear response
function is used) are
derived in Sec.~\ref{sec:working_eqs_ind_densities} for their
evaluation. Finally, we show in Sec.~\ref{sec:ind_KS_pots_from_ens_DFAs}
that, in the present context, state-specific KS potentials can 
emerge from well-identified ensemble density-functional
approximations, the exact Hx-only approximation (see Sec.~\ref{sec:ind_exact_Hx_approximation}) being one of them, the
second one consisting in recycling the regular ground-state Hxc
functional (see Sec.~\ref{sec:OO-DFT_excited-states}). In the latter case, the orbital-optimized DFT of excited states is actually recovered. Conclusions and
perspectives are given in Sec.~\ref{sec:conclusions}.

\section{Exact theory}

\subsection{Brief review of ensemble DFT}\label{sec:review_eDFT}

Unlike in regular DFT, where the ground-state electronic density is
determined from a single many-electron wavefunction (a so-called pure
state), densities are evaluated as
weighted sums of pure-state densities in the context of ensemble DFT.
Such an extension of DFT has various conceptual and practical
advantages~\cite{Teale2022_DFT_exchange,Cernatic2022}, one of them being the possibility to describe, in principle
exactly, electronic excitations. The present work deals with the
Theophilou--Gross--Oliveira--Kohn (TGOK) flavor of ensemble
DFT~\cite{JPC79_Theophilou_equi-ensembles,gross1988rayleigh, gross1988density,oliveira1988density},
where neutral excitations only are described. Note that charged
electronic excitations have been recently incorporated into the theory,
thus leading to a general (so-called extended $N$-centered) ensemble DFT of
electronic excited states~\cite{Cernatic2024_Neutral,cernatic2024extended_doubles}. In TGOK ensemble DFT, that
we simply refer to as ensemble DFT in the rest of the paper, the
quantity of interest is the ensemble energy,  
\be\label{eq:def_true_ens_ener}
E^{\bxi}:=\sum_{\nu\geq 0}\xi_\nu\,E_\nu,
\ee
which is a weighted sum of ground- ($\nu=0$) and excited-state ($\nu>0$)
$N$-electron energies. These energies are exact solutions to the $N$-electron Schr\"{o}dinger equation,  
\be\label{eq:phys_SE}
\hat{H}\myket{\Psi_\nu}\underset{\nu\geq 0}{=}E_\nu\myket{\Psi_\nu},
\ee
where the true physical electronic Hamiltonian operator,
\be\label{eq:phys_Hamil}
\hat{H}=\hat{T}+\hat{W}_{\rm ee}+\hat{V}_{\rm ext},
\ee
consists of the kinetic energy operator $\hat{T}$, the electron-electron
repulsion energy operator $\hat{W}_{\rm ee}$, and the external (nuclear attraction in standard quantum chemical calculations)
potential operator $\hat{V}_{\rm ext}=\int d\br\, v_{\rm
ext}(\br)\,\hat{n}(\br)$, where $\hat{n}(\br)$ is the electron density operator at
position $\br$ and $v_{\rm ext}: \br\mapsto v_{\rm ext}(\br)$ is the local external
potential. For the sake of simplicity and clarity, we will assume that
the ground state is not degenerate. This is by no means a restriction in
the theory, which can also tackle ground- and excited-state
multiplets~\cite{gross1988density}. If each ensemble weight $\xi_\nu$ (assigned to the
$\nu$th solution) fulfills the following ordering condition,
\be\label{eq:GOK_weights_ordering}
0\leq \xi_{\nu+1}\leq \xi_\nu,\;\;\nu\geq 0,
\ee  
then the ensemble energy can be determined variationally from a density
functional as follows,
\be\label{eq:ens_dens_VP}
E^{\bxi}=\min_n E^{\bxi}[n]=E^{\bxi}[n^{\bxi}],
\ee
where 
\be\label{eq:int_ens_ener_func_with_HK}
E^{\bxi}[n]=F^{\bxi}[n]+\left(v_{\rm ext}\vert n\right),
\ee
$\left(v_{\rm ext}\vert n\right):=\int
d\br\,v_{\rm ext}(\br)\,n(\br)$, and $F^{\bxi}[n]$ is the analog for
ensembles of the Hohenberg--Kohn (HK) functional~\cite{hktheo,gross1988density}, \ie, 
\be\label{eq:ens_HK_func_def}
F^{\bxi}[n]=\sum_{\nu\geq 0}\xi_\nu\,F^{{\bxi}}_{\nu}[n],
\ee
where 
\be\label{eq:ens_HK_func_ind_state_def}
F^{{\bxi}}_{\nu}[n]=\mybra{\Psi^{{\bxi}}_\nu[n]}\hat{T}+\hat{W}_{\rm
ee}\myket{\Psi^{{\bxi}}_\nu[n]}.
\ee
Note that, in Eq.~(\ref{eq:ens_dens_VP}), the ensemble weight values (which are
collected in $\bxi$) are arbitrarily {\it fixed}, up to the ordering
constraints of Eq.~(\ref{eq:GOK_weights_ordering}). For the trial
ensemble density $n: \br\mapsto n(\br)$, the ensemble HK functional is evaluated from the (orthonormalized) solutions $\Psi^{{\bxi}}_\nu[n]$ to the following ensemble
density-functional $N$-electron Schr\"{o}dinger equation,   
\be\label{eq:interacting_ens_dens_SE}
\left[\hat{T}+\hat{W}_{\rm
ee}+\hat{V}^{\bxi}[n]\right]\myket{\Psi^{{\bxi}}_\nu[n]}\underset{\nu\geq
0}{=}\check{E}^{\bxi}_\nu[n]\myket{\Psi^{{\bxi}}_\nu[n]},
\ee
where the local potential operator
\be\label{eq:many_elec_int_pot_from_local_pot}
\hat{V}^{\bxi}[n]:=\int d\br\, v^{\bxi}[n](\br)\,\hat{n}(\br)
\ee
is determined such that the ensemble density constraint,
\be\label{eq:int_ens_dens_constraint}
\sum_{\nu\geq 0}\xi_\nu\,n_{\Psi^{{\bxi}}_\nu[n]}=n,
\ee
is fulfilled. In Eq.~(\ref{eq:int_ens_dens_constraint}), 
\be
n_{\Psi}: \br\mapsto n_{\Psi}(\br)=\innerop{\Psi}{\hat{n}(\br)}{\Psi}
\ee
denotes the density of the normalized electronic wavefunction $\Psi$. 
If we want the ensemble density to integrate to
the (integer) number $N$ of electrons in the system under study, the
weights have to sum up to 1, according to Eq.~(\ref{eq:int_ens_dens_constraint}):
\be\label{eq:weights_sum_up_to_1}
\sum_{\nu\geq 0}\xi_\nu=1.
\ee
This implies that the weight assigned to the ground state is in fact determined from
those of the excited states, \ie, 
\be
\xi_0=1-\sum_{\lambda>0}\xi_\lambda,
\ee
where, from now on, $\lambda$ will be used as an index for excited
states. Consequently, the collection of {\it independent} weights $\bxi$ on which the
ensemble energy depends (see Eq.~(\ref{eq:def_true_ens_ener})) can be reduced to the
excited-state weights~\cite{deur2019ground},
\be
\bxi\equiv \left\{\xi_\lambda\right\}_{\lambda>0},
\ee
and the ensemble energy can be rewritten as follows, 
\be\label{eq:true_phys_ens_energy_exp_in_terms_xi_lambda}
E^{\bxi}=E_0+\sum_{\lambda>0}\xi_\lambda\left(E_\lambda-E_0\right).
\ee
Returning to the variational principle of Eq.~(\ref{eq:ens_dens_VP}),
the minimizing density $n^{\bxi}$ is the true physical ensemble density,
\ie, the weighted sum of the exact ground- and excited-state densities:
\begin{subequations}
\begin{align}
\label{eq:phys_ens_dens}
n^{\bxi}
&=\sum_{\nu\geq 0}\xi_\nu\,n_{\Psi_\nu}
\\
\label{eq:phys_ens_dens_exp_EX_weights}
&=n_{\Psi_0}+\sum_{\lambda>0}\xi_\lambda\left(n_{\Psi_\lambda}-n_{\Psi_0}\right).
\end{align}
\end{subequations}
As a final comment about the ensemble HK functional $F^{\bxi}[n]$, we note that, according to Eqs.~(\ref{eq:ens_HK_func_def}),
(\ref{eq:ens_HK_func_ind_state_def}),
(\ref{eq:interacting_ens_dens_SE}), and
(\ref{eq:many_elec_int_pot_from_local_pot}), any infinitesimal variation
$n\rightarrow n+\delta n$ of the ensemble density (with $\int d\br
\,\delta n(\br)=0$, to preserve the number of electrons) induces the following
variation, 
\begin{subequations}\label{eq:delta_F_ens}
\begin{align}
\delta F^{{\bxi}}[n]=\sum_{\nu\geq
0}\xi_\nu
&
\Bigg(\check{E}^{\bxi}_\nu[n]\delta\left\{\inner{\Psi^{{\bxi}}_\nu[n]}{\Psi^{{\bxi}}_\nu[n]}\right\}
\\
&\quad
-\int d\br\, v^{\bxi}[n](\br)\delta n_{\Psi^{{\bxi}}_\nu[n]}(\br) 
\Bigg),
\end{align}
\end{subequations}
thus leading, according to the ensemble density constraint of
Eq.~(\ref{eq:int_ens_dens_constraint}), to the key relation
\be\label{eq:int_ens_dens_func_pot_as_derivative}
v^{\bxi}[n]\equiv-\dfrac{\delta F^{\bxi}[n]}{\delta n} 
\ee
that will be exploited later in the paper.\\

Like in regular ground-state DFT, the commonly used KS formulation of
ensemble DFT~\cite{gross1988density,Cernatic2022} is obtained by considering the noninteracting (kinetic
energy) analog of $F^{{\bxi}}[n]$,
\be\label{eq:ens_Ts_func_def}
T_{\rm s}^{\bxi}[n]=\sum_{\nu\geq 0}\xi_\nu\,T^{{\bxi}}_{{\rm
s},\nu}[n],
\ee
with
\be\label{eq:ens_Ts_func_ind_state_def}
T^{{\bxi}}_{{\rm
s},\nu}[n]=\mybra{\Phi^{{\bxi}}_\nu[n]}\hat{T}\myket{\Phi^{{\bxi}}_\nu[n]},
\ee
where the orthonormalized ensemble density-functional KS wavefunctions $\Phi^{{\bxi}}_\nu[n]$ are solutions
to the $N$-electron KS equation
\be\label{eq:ens_dens_KSE}
\left[\hat{T}
+\hat{V}_{\rm s}^{\bxi}[n]\right]\myket{\Phi^{{\bxi}}_\nu[n]}\underset{\nu\geq
0}{=}\check{\mathcal{E}}^{\bxi}_\nu[n]\myket{\Phi^{{\bxi}}_\nu[n]},
\ee
the local KS potential operator
\be\label{eq:many_elec_dens_KS_pot_from_local_KS_pot}
\hat{V}_{\rm s}^{\bxi}[n]=\int d\br\, v_{\rm s}^{\bxi}[n](\br)\,\hat{n}(\br)
\ee
ensuring that the noninteracting KS ensemble reproduces the density $n$,
\ie, 
\begin{subequations}\label{eq:ens_KS_dens_constraint}
\begin{align}
n&=\sum_{\nu\geq 0}\xi_\nu n_{\Phi^{{\bxi}}_\nu[n]}
\\
\label{eq:ens_KS_dens_constraint_sum_lambda}
&=n_{\Phi^{{\bxi}}_0}[n]+\sum_{\lambda>0}\xi_\lambda
\left(n_{\Phi^{{\bxi}}_\lambda[n]}-n_{\Phi^{{\bxi}}_0[n]}\right).
\end{align}
\end{subequations}
Let us stress that, as further discussed in
Sec.~\ref{sec:ind_exact_Hx_approximation}, the KS wavefunctions are not
necessarily Slater determinants~\cite{gould2017hartree}. They are 
more generally configuration state functions. Once the analog for ensembles of the Hxc
density-functional energy, 
\be\label{eq:ens_KS_decomp}
E^{\bxi}_{\rm Hxc}[n]=F^{\bxi}[n]-T_{\rm s}^{\bxi}[n],
\ee
which is by construction {\it weight-dependent}, has been
introduced, it becomes possible to evaluate {\`{a}} la KS and in
principle exactly the ensemble energy (see Eqs.~(\ref{eq:ens_dens_VP})
and (\ref{eq:int_ens_ener_func_with_HK})):
\begin{subequations}
\begin{align}
E^{\bxi}&=\min_n\left\{T_{\rm s}^{\bxi}[n]+E^{\bxi}_{\rm
Hxc}[n]+\contract{v_{\rm ext}}{n}\right\} 
\\
&=\sum_{\nu\geq 0}\xi_\nu\innerop{\Phi^{\bxi}_\nu}{\hat{T}+\hat{V}_{\rm
ext}}{\Phi^{\bxi}_\nu}+E^{\bxi}_{\rm
Hxc}[n^{\bxi}],
\end{align}
\end{subequations}
where, according to Eq.~(\ref{eq:ens_KS_dens_constraint}), the minimizing ensemble density-functional KS wavefunctions
\be\label{eq:exact_KS_wfs}
\Phi^{\bxi}_\nu:=\Phi^{\bxi}_\nu[n^{\bxi}]
\ee
reproduce the true physical ensemble density, \ie,
\be\label{eq:exact_ens_KS_mapping}
\sum_{\nu\geq 0}\xi_\nu
n_{\Psi_\nu}=n^{\bxi}=n^{\bxi}_{\rm s}:=\sum_{\nu\geq 0}\xi_\nu
n_{\Phi^{\bxi}_\nu}.
\ee
Note that, according to the ensemble density-functional KS Eq.~(\ref{eq:ens_dens_KSE}) and by analogy with Eq.~(\ref{eq:delta_F_ens}),
\begin{subequations}\label{eq:delta_Ts_ens}
\begin{align}
\delta T_{\rm s}^{\bxi}[n]=\sum_{\nu\geq
0}\xi_\nu
&
\Bigg(\check{\mathcal{E}}^{\bxi}_\nu[n]\delta\left\{\inner{\Phi^{{\bxi}}_\nu[n]}{\Phi^{{\bxi}}_\nu[n]}\right\}
\\
&\quad
-\int d\br\, v_{\rm s}^{\bxi}[n](\br)\delta
n_{\Phi^{{\bxi}}_\nu[n]}(\br) 
\Bigg),
\end{align}
\end{subequations}
thus leading to the expected ensemble density-functional KS potential
expression (see Eqs.~(\ref{eq:int_ens_dens_func_pot_as_derivative}),
(\ref{eq:ens_KS_dens_constraint}) and (\ref{eq:ens_KS_decomp})),
\be\label{eq:KS_ens_dens_pot_exp}
v_{\rm s}^{\bxi}[n]=-\dfrac{\delta T_{\rm s}^{\bxi}[n]}{\delta
n}=v^{\bxi}[n]+v^{\bxi}_{\rm Hxc}[n],
\ee
where $v^{\bxi}_{\rm Hxc}[n]={\delta E^{\bxi}_{\rm Hxc}[n]}/{\delta n}$
is the ensemble density-functional Hxc potential. As a result, the exact KS
potential, from which the true physical ensemble density can be
reproduced at the noninteracting level of calculation, simply reads   
\be\label{eq:true_ens_KS_pot}
v_{\rm s}^{{\bxi}}:=\left.v_{\rm
s}^{{\bxi}}[n]\right|_{n=n^{\bxi}}=v_{\rm ext}+v^{\bxi}_{\rm
Hxc}[n^{\bxi}],
\ee
since, according to Eqs.~(\ref{eq:phys_SE}), (\ref{eq:phys_Hamil}), and (\ref{eq:phys_ens_dens}),
\be\label{eq:int_pot_equal_ext_pot}
v^{\bxi}[n^{\bxi}]=v_{\rm ext}.
\ee

\subsection{Ensemble density-functional stationarity of ground and
excited energy
levels}\label{sec:ens_dens_func_energy_levels_and_stat_cond}

It is clear from the previous section how an ensemble
energy can be evaluated in a DFT way. What is less clear is how each
component of the ensemble energy
(\ie, the different energy levels $E_\nu$), which is in fact the quantity of
interest, can be determined {\it individually}
within the present ensemble DFT formalism. The evaluation of the energy
levels {\it on top} of a single self-consistent ensemble KS-DFT calculation
(as described by Eqs.~(\ref{eq:ens_dens_KSE}), (\ref{eq:exact_KS_wfs}),
(\ref{eq:exact_ens_KS_mapping}), and (\ref{eq:true_ens_KS_pot})), has
already been discussed in Ref.~\citenum{deur2019ground}. What we are
aiming at here is different. We would like to derive a DFT for a
specific energy level whose construction fully relies on the ensemble
density. This question has been addressed recently by Gould from a
different perspective, namely that of ensemble potential functional
theory~\cite{gould2024stationaryconditionsexcitedstates}. We will follow
a different path and derive, instead, an ensemble DFT of energy levels
(where the ensemble density is the sole basic variable).
As shown later in Sec.~\ref{sec:OO-DFT_excited-states}, the present
formalism will enable us to clearly identify the ensemble
density-functional approximations underlying practical orbital-optimized DFT computations of
excited states~\cite{Levi20_Variational,Ivanov21_Method,Hait21_Orbital,Schmerwitz22_Variational}.\\    

For our purpose, let us first introduce the following
decomposition of the to-be-minimized ensemble density-functional energy
(see Eqs.~(\ref{eq:int_ens_ener_func_with_HK}),
(\ref{eq:ens_HK_func_def}), and (\ref{eq:int_ens_dens_constraint})),
\be
E^{\bxi}[n]=\sum_{\nu\geq 0}\xi_\nu
E^{\bxi}_\nu[n],
\ee
where, according to Eq.~(\ref{eq:ens_HK_func_ind_state_def}), the
$\nu$th
ensemble density-functional energy level reads 
\begin{subequations}
\begin{align}
\label{eq:ind_nu_energy_dens_func}
E^{\bxi}_{\nu}[n]&=F_\nu^{\bxi}[n]+\left(v_{\rm ext}\vert
n_{\Psi^{\bxi}_\nu[n]}\right)
\\
\label{eq:ind_state_func_from_true_Hamilt}
&=\innerop{\Psi^{\bxi}_\nu[n]}{\hat{H}}{\Psi^{\bxi}_\nu[n]}.
\end{align}
\end{subequations}
The exact $\nu$th energy level is recovered when $n$ equals the exact
ensemble density, \ie,
\be\label{eq:ind_nu_energy_dens_func_for_exact_ens_dens}
E^{\bxi}_{\nu}[n^{\bxi}]=E_\nu,
\ee
since, according to Eq.~(\ref{eq:int_pot_equal_ext_pot}),
\be\label{eq:exact_true_physical_wfs}
\Psi^{\bxi}_\nu[n^{\bxi}]=\Psi_\nu.
\ee 
As readily seen from
Eq.~(\ref{eq:ind_nu_energy_dens_func_for_exact_ens_dens}), in the
present context, ground- and
excited-state energy levels are evaluated as functionals of the ensemble
density. If we adopt a state-specific perspective, it means that,
as we explore the landscape of ensemble densities $n$
for fixed ensemble weight values $\bxi$, we can in principle reach any energy level without having to evaluate the full
ensemble energy, which is of course very appealing for practical
purposes. In order to determine that specific level, we need a
stationarity condition. While the one fulfilled by the ensemble
density-functional energy trivially follows from the variational
principle of Eq.~(\ref{eq:ens_dens_VP}), \ie,         
\be
\left.\delta E^{\bxi}[n]\right|_{n=n^{\bxi}}=0,
\ee
it is less obvious for the excited-state energy levels because they are not local
minima of the energy. But they are stationary points of the energy,
which makes them also stationary when expressed as functionals of the
ensemble density. Indeed, according to Eqs.~(\ref{eq:phys_SE}),
(\ref{eq:ind_state_func_from_true_Hamilt}), and (\ref{eq:exact_true_physical_wfs}), 
\begin{subequations}\label{eq:ind_state_within_ens_VP}
\begin{align}
\left.\delta E^{\bxi}_{\nu}[n]\right|_{n=n^{\bxi}}
&=2\left.\mybra{\delta
\Psi^{\bxi}_\nu[n]}\hat{H}\myket{\Psi_\nu}\right|_{n=n^{\bxi}}
\\
&=E_\nu \left.\delta
\left\{\inner{\Psi^{\bxi}_\nu[n]}{\Psi^{\bxi}_\nu[n]}\right\}\right|_{n=n^{\bxi}} 
\\
&=0,
\end{align}
\end{subequations}
where real algebra has been used, for simplicity.
Eq.~(\ref{eq:ind_state_within_ens_VP}) echoes Eq.~(12) of
Ref.~\citenum{gould2024stationaryconditionsexcitedstates}, where
infinitesimal variations of the local potential are
considered instead.\\

The noninteracting analog of Eq.~(\ref{eq:ind_state_within_ens_VP}),
which will be exploited later (in
Eq.~(\ref{eq:final_ind_state_stat_cond})), can
be obtained by introducing the following ensemble density-functional KS
energy level (see Eq.~(\ref{eq:true_ens_KS_pot})),   
\begin{subequations}
\begin{align}
\label{eq:ind_KS_ener_fun_for_exact_vs}
\mathcal{E}_{\nu}^{\bxi}[n]&=T_{{\rm s},\nu}^{\bxi}[n]+\contract{v_{\rm s}^{{\bxi}}
}{n_{\Phi^{{\bxi}}_\nu[n]}}
\\
&=\innerop{\Phi^{{\bxi}}_\nu[n]}{\hat{T}+\hat{V}^{\bxi}_{\rm
s}}{\Phi^{{\bxi}}_\nu[n]},
\end{align}
\end{subequations}
where $\hat{V}^{\bxi}_{\rm s}=\int d\br\, v^{\bxi}_{\rm
s}(\br)\,\hat{n}(\br)$ is the exact KS potential operator,
so that, according to the ensemble KS Eq.~(\ref{eq:ens_dens_KSE}),
\begin{subequations}\label{eq:ind_KS_state_within_ens_VP}
\begin{align}
\label{eq:ens_KS_stat_cond_line1}
\left.\delta \mathcal{E}_{\nu}^{\bxi}[n]\right|_{n=n^{\bxi}}
&
=2\left.\innerop{\delta {\Phi^{{\bxi}}_\nu[n]}}{\hat{T}+\hat{V}^{\bxi}_{\rm
s}}{\Phi^{\bxi}_{\nu}}\right|_{n=n^{\bxi}}
\\
&={\mathcal{E}}^{\bxi}_\nu\left.\delta
\left\{\inner{\Phi^{\bxi}_\nu[n]}{\Phi^{\bxi}_\nu[n]}\right\}\right|_{n=n^{\bxi}} 
\\
&=0,
\end{align}
\end{subequations}
where we used the shorthand notation (see Eqs.~(\ref{eq:ens_dens_KSE}) and (\ref{eq:true_ens_KS_pot}))
\be\label{eq:notation_fictitious_KS_ener}
{\mathcal{E}}^{\bxi}_\nu:=\mathcal{E}_{\nu}^{\bxi}[n^{\bxi}]=\check{\mathcal{E}}^{\bxi}_\nu[n^{\bxi}].
\ee

As discussed in further details in Sec.~\ref{sec:ind_state_KSDFT}, the
above KS stationarity condition is not sufficient for establishing a DFT of
excited-state energy levels, which is the purpose of the present work.
First of all, the auxiliary KS energies
$\left\{{\mathcal{E}}^{\bxi}_\nu\right\}_{\nu\geq 0}$ are
{\it not} the true physical energies~\cite{deur2019ground}. Secondly, in ensemble DFT, the KS
ensemble reproduces the physical ensemble density, not the individual
physical densities. The concept of density-driven correlation emerges
from that
observation~\cite{PRL19_Gould_DD_correlation,Fromager_2020,Cernatic2022,gould2024stationaryconditionsexcitedstates}.       
Therefore, in order to set up an exact ensemble DFT of
ground and excited energy levels, we should start from the HK-type
ensemble density-functional energy level expression of
Eq.~(\ref{eq:ind_nu_energy_dens_func}) and insert the following
individual-state KS decomposition,
\be\label{eq:KS_decomp_Hxc_energy_level}
F_\nu^{\bxi}[n]=T_{{\rm s},\nu}^{\bxi}[n]+{E}^{\bxi}_{{\rm Hxc},\nu}[n],
\ee
where, according to Eqs.~(\ref{eq:ens_HK_func_ind_state_def}) and
(\ref{eq:ens_Ts_func_ind_state_def}), the $\nu$th component of the
ensemble Hxc functional reads
\be
\begin{aligned}
{E}^{\bxi}_{{\rm Hxc},\nu}[n]&=\mybra{\Psi^{{\bxi}}_\nu[n]}\hat{T}+\hat{W}_{\rm
ee}\myket{\Psi^{{\bxi}}_\nu[n]}
\\
&\quad-\mybra{\Phi^{{\bxi}}_\nu[n]}\hat{T}\myket{\Phi^{{\bxi}}_\nu[n]}.
\end{aligned}
\ee
This leads to the $\nu$th energy level ensemble-based KS expression, 
\be\label{eq:energy_level_ens_KS_exp}
E^{\bxi}_\nu[n]=T_{{\rm s},\nu}^{\bxi}[n]+{E}^{\bxi}_{{\rm
Hxc},\nu}[n]+\contract{v_{\rm ext}}{n_{\Psi^{\bxi}_\nu[n]}},
\ee
which, as readily seen, involves the true density of the $\nu$th state
(last term on the right-hand side of Eq.~(\ref{eq:energy_level_ens_KS_exp})),
and the resulting stationarity condition  
\be\label{eq:ens_based_energy_level_KS-decomp_stat_cond}
0=\left.\delta \left\{T_{{\rm s},\nu}^{\bxi}[n]+{E}^{\bxi}_{{\rm Hxc},\nu}[n]+\left(v_{\rm ext}\vert
n_{\Psi^{\bxi}_\nu[n]}\right)\right\}\right|_{n=n^{\bxi}}.
\ee
What might be unclear in this construction is how the individual Hxc ensemble
density functional ${E}^{\bxi}_{{\rm Hxc},\nu}[n]$ is related to the original ensemble one
$E^{\bxi}_{\rm Hxc}[n]$ (see Eq.~(\ref{eq:ens_KS_decomp})). The explicit
answer (see the proof in Appendix~\ref{sec:ind_from_ens_func}) is given below,   
\be
\label{eq:ind_state_Hxc_func_from_ens_func}
\begin{aligned}
&E^{\bxi}_{{\rm Hxc},\nu}[n]=
E^{\bxi}_{\rm Hxc}[n]
+\sum_{\lambda>0}\left(\delta_{\lambda\nu}-\xi_\lambda\right)\dfrac{\partial E^{\bxi}_{\rm Hxc}[n]}{\partial
\xi_\lambda}
\\
&\quad-\contract{\dfrac{\delta E^{\bxi}_{\rm Hxc}[n]}{\delta
n}}{n}+\contract{\dfrac{\delta F^{{\bxi}}[n]}{\delta
n}}{n_{\Psi^{{\bxi}}_\nu[n]}}
\\
&\quad-\contract{\dfrac{\delta T_{\rm s}^{{\bxi}}[n]}{\delta
n}}{n_{\Phi^{{\bxi}}_\nu[n]}},
\end{aligned}
\ee
and it fully relies on the fact that, for a fixed local potential, the
ensemble energy varies {\it linearly}
with the ensemble weights (see Eq.~(\ref{eq:true_phys_ens_energy_exp_in_terms_xi_lambda})).
Eq.~(\ref{eq:ind_state_Hxc_func_from_ens_func}) enables to retrieve, in principle exactly, any ensemble
density-functional energy level $\nu$ (see Eq.~(\ref{eq:energy_level_ens_KS_exp})) from the $\nu$th component of the ensemble non-interacting
kinetic energy functional (see Eq.~(\ref{eq:ens_Ts_func_def})) as follows, 
\be\label{eq:ind_ener_ens_func}
\begin{aligned}
&E^{\bxi}_{\nu}[n]=T_{{\rm
s},\nu}^{\bxi}[n]+E^{\bxi}_{\rm Hxc}[n]
\\
&+\sum_{\lambda>0}\left(\delta_{\lambda\nu}-\xi_\lambda\right)\dfrac{\partial E^{\bxi}_{\rm Hxc}[n]}{\partial
\xi_\lambda}
-\contract{\dfrac{\delta E^{\bxi}_{\rm Hxc}[n]}{\delta
n}}{n}
\\
&+\contract{\dfrac{\delta F^{{\bxi}}[n]}{\delta
n}+v_{\rm ext}}{n_{\Psi^{{\bxi}}_\nu[n]}}
-\contract{\dfrac{\delta T_{\rm s}^{{\bxi}}[n]}{\delta
n}}{n_{\Phi^{{\bxi}}_\nu[n]}}.
\end{aligned}
\ee
Eqs.~(\ref{eq:ind_state_Hxc_func_from_ens_func}) and (\ref{eq:ind_ener_ens_func}) are the first key
result of this work. Eq.~(\ref{eq:ind_ener_ens_func}) generalizes to any ensemble
density $n$ the exact energy level expression of Eq.~(10) in
Ref.~\citenum{Fromager_2020},
which is recovered when $n$ equals the true physical ensemble density
$n^{\bxi}$, thus allowing for the derivation of a stationarity condition
for each individual state, as shown in the following. Note that, since
the interacting and KS individual densities do not match, {\it a priori}, we do {\it not} reach in
Eq.~(\ref{eq:ind_state_Hxc_func_from_ens_func}) an expression that is exclusively written
in terms of the 
ensemble Hxc functional and its derivatives (see the last two terms on the right-hand side). 
In the rest of the paper, we will decipher the stationarity condition of Eq.~(\ref{eq:ens_based_energy_level_KS-decomp_stat_cond}), on the basis of
Eq.~(\ref{eq:ind_state_Hxc_func_from_ens_func}), and then construct
approximate formulations that could be
used, in future works, to design alternative (\ie, more state-specific) KS decompositions to that
of Eq.~(\ref{eq:KS_decomp_Hxc_energy_level}), which relies on the ensemble
KS states. 

\subsection{Deciphering the stationarity of energy levels when expressed
in ensemble Kohn--Sham DFT}\label{sec:ind_state_KSDFT}

While the exact evaluation of the $\nu$th energy
level ($\nu\geq 0$) from a KS-like functional of the ensemble density has been made more
explicit in Eqs.~(\ref{eq:ind_state_Hxc_func_from_ens_func}) and
(\ref{eq:ind_ener_ens_func}), it is still unclear what the implications
of the resulting stationarity
condition (see
Eq.~(\ref{eq:ens_based_energy_level_KS-decomp_stat_cond})) are. For
example, one may wonder if it ultimately leads to an
individual-state KS-like equation. In order to address this question,
which is the purpose of this section, let us first insert
Eq.~(\ref{eq:ind_state_Hxc_func_from_ens_func}) into
Eq.~(\ref{eq:ens_based_energy_level_KS-decomp_stat_cond}), or,
equivalently, evaluate the density functional derivative of
Eq.~(\ref{eq:ind_ener_ens_func}) at the physical ensemble density
$n=n^{\bxi}$. According to Eqs.~(\ref{eq:int_ens_dens_func_pot_as_derivative}),
(\ref{eq:int_pot_equal_ext_pot}), and (\ref{eq:KS_ens_dens_pot_exp}), we obtain  
the following (more explicit) stationarity condition, 
\be\label{eq:explicit_ener_level_KS_exp}
\begin{aligned}
&\Bigg[\delta T_{{\rm
s},\nu}^{\bxi}[n]+\delta E^{\bxi}_{\rm Hxc}[n]
\\
&
+\sum_{\lambda>0}\left(\delta_{\lambda\nu}-\xi_\lambda\right)\delta
\left\{\dfrac{\partial E^{\bxi}_{\rm Hxc}[n]}{\partial
\xi_\lambda}\right\}
\\
&
-\contract{\dfrac{\delta E^{\bxi}_{\rm Hxc}[n]}{\delta                                      
n}}{\delta n} 
-\contract{\delta n}{f^{\bxi}_{\rm Hxc}[n]
\star n}
\\
&
-\contract{\delta n}{\dfrac{\delta v^{{\bxi}}[n]}{\delta
n}\star n_{\Psi^{{\bxi}}_\nu[n]}}
+\contract{\delta n}{\dfrac{\delta v_{\rm s}^{{\bxi}}[n]}{\delta
n}\star n_{\Phi^{{\bxi}}_\nu[n]}}
\\
&
+\contract{v_{\rm s}^{{\bxi}}[n]
}{\delta n_{\Phi^{{\bxi}}_\nu[n]}}\Bigg]_{n=n^{\bxi}}=0,
\end{aligned}
\ee
where $f^{\bxi}_{\rm Hxc}[n]={\delta v_{\rm Hxc}^{\bxi}[n]}/{\delta n}$
is the ensemble density-functional Hxc kernel, \ie,
\be\label{eq:ens_Hxc_kernel_func}
\begin{aligned}
f^{\bxi}_{\rm Hxc}[n]&: (\br',\br)\mapsto \dfrac{\delta v_{\rm Hxc}^{\bxi}[n](\br)}{\delta n (\br')}
=\dfrac{\delta^2 E^{\bxi}_{\rm Hxc}[n]}{\delta
n(\br')\delta n(\br)},
\end{aligned}
\ee
and the shorthand notation 
\be
f\star n: \br'\mapsto \int d\br\,f(\br',\br)\,n(\br)
\ee
has been used. After some minor simplifications (see also
Eqs.~(\ref{eq:exact_KS_wfs}) and (\ref{eq:exact_true_physical_wfs})), Eq.~(\ref{eq:explicit_ener_level_KS_exp}) becomes   
\be\label{eq:final_ind_state_stat_cond}
\begin{aligned}
&\Bigg[\delta T_{{\rm
s},\nu}^{\bxi}[n]
+\contract{v_{\rm s}^{{\bxi}}
}{\delta n_{\Phi^{{\bxi}}_\nu[n]}}
\Bigg]_{n=n^{\bxi}}
\\
&
+
\Bigg[\sum_{\lambda>0}\left(\delta_{\lambda\nu}-\xi_\lambda\right)
\contract{\delta n}{\dfrac{\partial v^{\bxi}_{\rm Hxc}[n]}{\partial \xi_\lambda}}
\\
&
\quad\quad
-\contract{\delta n}{f^{\bxi}_{\rm Hxc}
\star n^{\bxi}}
-\contract{\delta n}{\dfrac{\delta v^{{\bxi}}[n]}{\delta
n}\star n_{\Psi_\nu}}
\\
&
\quad\quad
+\contract{\delta n}{\dfrac{\delta v_{\rm s}^{{\bxi}}[n]}{\delta
n}\star n_{\Phi^{{\bxi}}_\nu}}
\Bigg]_{n=n^{\bxi}}
=0,
\end{aligned}
\ee
where the first and last terms on the left-hand side of Eq.~(\ref{eq:explicit_ener_level_KS_exp})
have been combined (see also Eq.~(\ref{eq:true_ens_KS_pot})), and
\be\label{eq:short-hand_notation_exact_ens_Hxc_kernel}
f^{\bxi}_{\rm Hxc}:=f^{\bxi}_{\rm Hxc}[n^{\bxi}].
\ee
With some additional
derivations, which are presented in
Appendix~\ref{sec:appendix_simpl_stat_cond}, and
Eq.~(\ref{eq:ind_KS_ener_fun_for_exact_vs}), we finally obtain the
substantially simplified and
compact stationarity condition for each energy level, as it should read
in ensemble KS-DFT, 
\begin{subequations}\label{eq:stat_cond_true_from_KS}
\begin{align}
\label{eq:KS_ens_stat_cond_part_stat_cond}
&\left.
\delta \mathcal{E}_{\nu}^{\bxi}[n]
\right|_{n=n^{\bxi}}
\\
&
\label{eq:KS_rsp_func_part_stat_cond}
+\contract
{\delta n}
{
\sum_{\lambda>0}\left(\delta_{\lambda\nu}
-\xi_\lambda\right)\dfrac{\partial
v^{\bxi}_{\rm Hxc}}{\partial
\xi_\lambda}
-\left[\chi_{\rm s}^{\bxi}\right]^{-1}\star
\left(n_{\Psi_\nu}
-
n_{\Phi_\nu^{\bxi}}
\right)
}
\\
&
=0
.
\end{align}
\end{subequations}
In Eq.~(\ref{eq:KS_rsp_func_part_stat_cond}),
\be\label{eq:true_Hxc_pot_notation}
v^{\bxi}_{\rm Hxc}:=v^{\bxi}_{\rm Hxc}[n^{\bxi}]=v_{\rm
s}^{\bxi}-v_{\rm ext}
\ee
is the exact Hxc
potential from which the true ensemble density $n^{\bxi}$ can be
reproduced and
$\chi^{\bxi}_{\rm
s}:\;(\br,\br')\mapsto \chi^{\bxi}_{\rm
s}({\br},{\br'})$ is the static ensemble KS density-density linear
response function (see Eqs.~(\ref{eq:true_ens_KS_pot}) and (\ref{eq:exact_ens_KS_mapping})), \ie, 
\be\label{eq:ens_KS_state_dens-dens_rsp_func}
\chi^{\bxi}_{\rm
s}=\dfrac{\delta n^{\bxi}_{\rm s}}{\delta v^{\bxi}_{\rm s}}=\sum_{\kappa\geq 0}\xi_\kappa \chi^{\bxi}_{{\rm
s},\kappa},
\ee
where the summation in $\kappa\geq 0$ runs over ground and excited KS
states, and the individual-state KS linear response functions can be expressed
explicitly as follows (see Eqs.~(\ref{eq:ens_dens_KSE}), (\ref{eq:exact_KS_wfs}), and (\ref{eq:notation_fictitious_KS_ener})), 
\bse
\begin{align}
\label{eq:ind_KS_state_dens-dens_rsp_func}
\chi^{\bxi}_{{\rm
s},\kappa}({\br},{\br'})
&=\dfrac{\delta n_{\Phi^{\bxi}_\kappa}(\br)}{\delta v^{\bxi}_{\rm s}(\br')}
\\
&=2\sum_{0\leq\mu\neq
\kappa}\dfrac{\langle\Phi^{\bxi}_\kappa\vert\hat{n}({\br})\vert
\Phi^{\bxi}_\mu\rangle\langle\Phi^{\bxi}_\mu\vert\hat{n}({\br}')\vert
\Phi^{\bxi}_\kappa\rangle}{\mathcal{E}^{\bxi}_{\kappa}-\mathcal{E}^{\bxi}_{\mu}}
.
\end{align}
\ese
\\

Let us now interpret Eq.~(\ref{eq:stat_cond_true_from_KS}), which is our
second key result. As the present state-specific KS-DFT relies on a KS
ensemble, which is
characterized by the noninteracting stationarity condition of
Eq.~(\ref{eq:ind_KS_state_within_ens_VP}), the true physical 
stationarity condition of Eq.~(\ref{eq:stat_cond_true_from_KS}), which
holds for any ensemble density variation $\delta n$ and where
the term in Eq.~(\ref{eq:KS_ens_stat_cond_part_stat_cond}) cancels out,
leads to  
\be\label{eq:true_dens_from_stat_cond}
n_{\Psi_\nu}-n_{\Phi_\nu^{\bxi}}=\sum_{\lambda>0}\left(\delta_{\lambda\nu}
-\xi_\lambda\right)\chi_{\rm s}^{\bxi}\star\dfrac{\partial
v^{\bxi}_{\rm Hxc}}{\partial
\xi_\lambda}.
\ee
Eq.~(\ref{eq:true_dens_from_stat_cond}) simply tells us
how the true density $n_{\Psi_\nu}$ of the targeted $\nu$th physical state can
be retrieved from the KS ensemble. Thus, the theory
gives ultimately access, and in principle exactly, to the true energy
level $E_\nu$ (see
Eqs.~(\ref{eq:ind_nu_energy_dens_func}) and
(\ref{eq:ind_nu_energy_dens_func_for_exact_ens_dens})).\\

At this point, several comments should be made. Firstly, if we
(somehow arbitrarily) define the $\nu$th component of the ensemble Hxc
potential as follows,
simply by analogy with the extraction of individual densities from the
linear-in-$\bxi$ physical ensemble density (see
Eq.~(\ref{eq:ind_dens_extraction_linearity_appendix})), 
\be\label{eq:def_ind_Hxc_pot}
{v}^{\bxi}_{{\rm Hxc},\nu}\underset{\nu\geq 0}{:=}v^{\bxi}_{\rm Hxc}+\sum_{\lambda>0}\left(\delta_{\lambda\nu}
-\xi_\lambda\right)\dfrac{\partial
v^{\bxi}_{\rm Hxc}}{\partial \xi_\lambda},
\ee
where, by construction (see Eq.~(\ref{eq:weights_sum_up_to_1})),
\be
\sum_{\nu\geq 0}\xi_\nu\,{v}^{\bxi}_{{\rm
Hxc},\nu}
=v^{\bxi}_{\rm Hxc},
\ee
then Eq.~(\ref{eq:true_dens_from_stat_cond}) can be rewritten in a more
compact way as follows,
\be
n_{\Psi_\nu}-n_{\Phi_\nu^{\bxi}}=\chi_{\rm
s}^{\bxi}\star\left({v}^{\bxi}_{{\rm Hxc},\nu}-v^{\bxi}_{\rm
Hxc}\right),
\ee
thus leading to
\be\label{eq:exact_ind_Hxc_pot_when_ind_densities_match}
n_{\Phi_\nu^{\bxi}}=n_{\Psi_\nu}\Leftrightarrow v^{\bxi}_{\rm
Hxc}={v}^{\bxi}_{{\rm Hxc},\nu}. 
\ee
Even though the perfect match of individual densities as assumed in the
left-hand side of Eq.~(\ref{eq:exact_ind_Hxc_pot_when_ind_densities_match}), in addition to
the ensemble one (see Eq.~(\ref{eq:exact_ens_KS_mapping})), is unlikely in general~\cite{PRL19_Gould_DD_correlation,Fromager_2020,Cernatic2022}, it is
interesting to notice that, in case it happened to be true,
then the naive expression of
Eq.~(\ref{eq:def_ind_Hxc_pot}) would actually provide an exact Hxc potential for
the state of interest (because that potential would be equal to the ensemble
one in this case, according to Eq.~(\ref{eq:exact_ind_Hxc_pot_when_ind_densities_match})). 
Therefore, it could be used as an approximate individual-state
Hxc potential in the general case. This point, which echoes the recent findings of Gould in the context of
ensemble potential functional
theory~\cite{gould2024stationaryconditionsexcitedstates}, is further discussed in Sec.~\ref{sec:ind_exact_Hx_approximation} at the
exact Hx-only level of approximation.\\      

Secondly, since $\partial
v^{\bxi}_{\rm Hxc}/\partial \xi_\lambda=\partial v_{\rm
s}^{\bxi}/\partial \xi_\lambda$ (see
Eq.~(\ref{eq:true_Hxc_pot_notation})), the deviation in density between
the true and KS wavefunctions, as written in
Eq.~(\ref{eq:true_dens_from_stat_cond}), can be expressed differently,
simply by noticing that (see Eqs.~(\ref{eq:ens_KS_state_dens-dens_rsp_func}) and
(\ref{eq:ind_KS_state_dens-dens_rsp_func})) 
\begin{subequations}
\begin{align}
\chi_{\rm s}^{\bxi}\star\dfrac{\partial
v^{\bxi}_{\rm Hxc}}{\partial
\xi_\lambda}&=\dfrac{\delta n_{\rm s}^{\bxi}}{\delta v_{\rm s}^{\bxi}}\star \dfrac{\partial
v^{\bxi}_{\rm s}}{\partial
\xi_\lambda}
\\
&=\sum_{\kappa\geq 0}\xi_\kappa \dfrac{\delta n_{\Phi^{\bxi}_{\kappa}}}{\delta v_{\rm s}^{\bxi}}\star \dfrac{\partial
v^{\bxi}_{\rm s}}{\partial
\xi_\lambda}
\\
\label{eq:deriv_KS_ens_dens_KSwf_deriv_part}
&=\sum_{\kappa\geq 0}\xi_\kappa\dfrac{\partial n_{\Phi^{\bxi}_{\kappa}}}{\partial
\xi_\lambda}, 
\end{align}
\end{subequations} 
so that Eq.~(9) of Ref.~\citenum{Fromager_2020} is recovered:  
\be\label{eq:ind_state_dens_extraction_PRL}
n_{\Psi_\nu}-n_{\Phi_\nu^{\bxi}}=\sum_{\lambda>0}\sum_{\kappa\geq 0}\left(\delta_{\lambda\nu}
-\xi_\lambda\right)\xi_\kappa\dfrac{\partial n_{\Phi^{\bxi}_{\kappa}}}{\partial
\xi_\lambda}.
\ee
While Eq.~(\ref{eq:ind_state_dens_extraction_PRL}) is, in the present
work, a direct consequence of the fact that the $\nu$th ensemble
density-functional energy level is stationary at the ensemble density
$n^{\bxi}$, it was simply deduced in Ref.~\citenum{Fromager_2020} from
the linear variation of $n^{\bxi}
$ with respect to the ensemble weights, as depicted in
Eq.~(\ref{eq:ind_dens_extraction_linearity_appendix}), and its mapping
onto the KS ensemble (see Eq.~(\ref{eq:exact_ens_KS_mapping})). This
becomes even more clear when rewritting
Eq.~(\ref{eq:ind_state_dens_extraction_PRL}) as follows,
\begin{subequations}
\begin{align}
n_{\Psi_\nu}&=n_{\Phi_\nu^{\bxi}}
\\
&\quad+\sum_{\lambda>0}\left(\delta_{\lambda\nu}
-\xi_\lambda\right)\left(\dfrac{\partial n_{\rm s}^{\bxi}}{\partial \xi_\lambda}-
\left(n_{\Phi_\lambda^{\bxi}}-n_{\Phi_0^{\bxi}}\right)\right)
\\
&=n_{\rm s}^{\bxi}+\sum_{\lambda>0}\left(\delta_{\lambda\nu}
-\xi_\lambda\right)\dfrac{\partial n_{\rm s}^{\bxi}}{\partial
\xi_\lambda}
\\
\label{eq:ind_dens_from_ens_dens_phys_main_text}
&=n^{\bxi}+\sum_{\lambda>0}\left(\delta_{\lambda\nu}
-\xi_\lambda\right)\dfrac{\partial n^{\bxi}}{\partial
\xi_\lambda}.
\end{align}
\end{subequations} 

Finally, as pointed out in Ref.~\citenum{Fromager_2020} and readily seen both 
from 
Eq.~(\ref{eq:ind_dens_from_ens_dens_phys_main_text}) and its original
form (Eq.~(\ref{eq:true_dens_from_stat_cond}) in conjunction with Eq.~(\ref{eq:true_Hxc_pot_notation})),  we
need, in order to
extract the true physical
density $n_{\Psi_\nu}$ from the KS ensemble, to evaluate the response $\partial n^{\bxi}/\partial \xi_\lambda$ of
the ensemble density to
infinitesimal variations of the ensemble weights. So far, no proper
working equations have been derived for that purpose, even in
Ref.~\citenum{Fromager_2020}, where a tentative derivation 
is presented in the supplementary material. Unlike in the latter, where
emphasize was put only on the 
coupling between the linear responses of the ensemble KS orbitals, a
complete solution to the problem is provided in the next 
Sec.~\ref{sec:working_eqs_ind_densities}.

\subsection{Working equations for evaluating the exact individual
densities}\label{sec:working_eqs_ind_densities}

We derive in this section a working equation for $\partial
n^{\bxi}/\partial \xi_\lambda$ so that exact individual-state densities
can be evaluated, according to
Eq.~(\ref{eq:ind_dens_from_ens_dens_phys_main_text}). Starting from the exact
density mapping onto the KS ensemble (see
Eq.~(\ref{eq:exact_ens_KS_mapping})) and
Eq.~(\ref{eq:deriv_KS_ens_dens_KSwf_deriv_part}), it comes  
\be\label{eq:deriv_ens_dens_wrt_weight_step1}
\dfrac{\partial n^{\bxi}}{\partial
\xi_\lambda}
=
\dfrac{\partial n_{\rm s}^{\bxi}}{\partial
\xi_\lambda}
=\left(n_{\Phi_\lambda^{\bxi}}-n_{\Phi_0^{\bxi}}\right)+\chi^{\bxi}_{\rm
s}\star\dfrac{\partial v_{\rm Hxc}^{\bxi}}{\partial
\xi_\lambda},
\ee
where, according to Eqs.~(\ref{eq:ens_Hxc_kernel_func}), (\ref{eq:short-hand_notation_exact_ens_Hxc_kernel}), and (\ref{eq:true_Hxc_pot_notation}),
\be\label{eq:weight_deriv_true_ens_Hxc_pot}
\dfrac{\partial v_{\rm Hxc}^{\bxi}}{\partial
\xi_\lambda}=\left.\dfrac{\partial
v^{\bxi}_{\rm Hxc}[n]}{\partial
\xi_\lambda}\right|_{n=n^{\bxi}}+f^{\bxi}_{\rm Hxc}\star \dfrac{\partial n^{\bxi}}{\partial
\xi_\lambda}.
\ee
If we make a formal analogy with ensemble DFT for fractional electron
numbers and the extraction of Fukui functions~\cite{Hellgren12_Effect,kraisler2013piecewise},
we notice that, in the commonly used Perdew-Parr-Levy-Balduz (PPLB)
approach~\cite{perdew1982density} (one would differentiate with respect
to the number of electrons in this case), the first term on the right-hand side of
Eq.~(\ref{eq:weight_deriv_true_ens_Hxc_pot}) would vanish. This is simply due to the
fact that the (ground-state) Hxc
functional has no weight dependence~\cite{Cernatic2022}. It only varies with the density
(that integrates to a fractional number of electrons). Note that it would not be
the case anymore if the alternative $N$-centered ensemble
formalism~\cite{senjean2018unified,Cernatic2022} were adopted instead.\\

If we now gather in the following quantity the contributions that can be
immediately evaluated from the KS system, 
\be\label{eq:gathering_KS_terms_deriv_ens_dens_wrt_weight}
\Delta
n^{\bxi}_{{\rm s},\lambda}
=\left(n_{\Phi_\lambda^{\bxi}}-n_{\Phi_0^{\bxi}}\right)+\chi^{\bxi}_{\rm
s}\star \left.\dfrac{\partial
v^{\bxi}_{\rm Hxc}[n]}{\partial
\xi_\lambda}\right|_{n=n^{\bxi}}, 
\ee
then combining Eqs.~(\ref{eq:deriv_ens_dens_wrt_weight_step1}) and (\ref{eq:weight_deriv_true_ens_Hxc_pot}) leads to
\be\label{eq:sc_eq_dnxi_over_d_xi_lambda}
\dfrac{\partial n^{\bxi}}{\partial
\xi_\lambda}=
\Delta
n^{\bxi}_{{\rm s},\lambda}
+\chi^{\bxi}_{\rm
s}\star f^{\bxi}_{\rm Hxc}\star \dfrac{\partial n^{\bxi}}{\partial
\xi_\lambda},
\ee
or, equivalently,
\be\label{eq:deriv_ens_dens_wrt_weight_step2}
\dfrac{\partial n^{\bxi}}{\partial
\xi_\lambda}=\left[\hat{1}-\chi^{\bxi}_{\rm
s}\star f^{\bxi}_{\rm Hxc}\right]^{-1}\star 
\Delta
n^{\bxi}_{{\rm s},\lambda},
\ee
which echoes Eq.~(14) of Ref.~\citenum{Hellgren12_Effect} for the exact
evaluation of Fukui
functions~\cite{Cohen94_Electronic,Cohen95_Reactivity}. Following
Ref.~\citenum{Hellgren12_Effect}, we can alternatively express $\partial
n^{\bxi}/\partial \xi_\lambda$ directly in terms of the true physical ensemble
density-density linear response function $\chi^{\bxi}=\delta
n^{\bxi}/\delta v_{\rm ext}$. Indeed, from the ensemble
density-functional Dyson equation (see
Eqs.~(\ref{eq:KS_ens_dens_pot_exp}) and (\ref{eq:ens_Hxc_kernel_func})),
\begin{subequations}
\begin{align}
\left[\chi^{\bxi}[n]\right]^{-1}
&=\dfrac{\delta v^{{\bxi}}[n]}{\delta
n}=\dfrac{\delta}{\delta n}\left(v_{\rm s}^{\bxi}[n]-v_{\rm Hxc}^{\bxi}[n]\right)
\\\label{eq:dens_func_Dyson_eq}
&
=\left[\chi_{\rm
s}^{\bxi}[n]\right]^{-1}-f^{\bxi}_{\rm Hxc}[n],
\end{align}
\end{subequations}
which gives, for $n=n^{\bxi}$,
\be\label{eq:dyson_n_equal_nxi}
\left[\chi^{\bxi}\right]^{-1}=\left[\chi_{\rm
s}^{\bxi}\right]^{-1}-f^{\bxi}_{\rm Hxc},
\ee 
we can proceed with the following simplification, 
\be
\begin{aligned}
\hat{1}-\chi^{\bxi}_{\rm
s}\star f^{\bxi}_{\rm Hxc}&=\hat{1}-\chi^{\bxi}_{\rm
s}\star\left(\left[\chi_{\rm
s}^{\bxi}\right]^{-1}-\left[\chi^{\bxi}\right]^{-1}\right)
\\
&=\chi^{\bxi}_{\rm
s}\star\left[\chi^{\bxi}\right]^{-1}.
\end{aligned}
\ee
Consequently, Eq.~(\ref{eq:deriv_ens_dens_wrt_weight_step2}) becomes
\be
\dfrac{\partial n^{\bxi}}{\partial
\xi_\lambda}=\chi^{\bxi}\star \left[\chi_{\rm
s}^{\bxi}\right]^{-1}\star 
\Delta
n^{\bxi}_{{\rm s},\lambda}
, 
\ee
thus leading, according to the Dyson Eq.~(\ref{eq:dyson_n_equal_nxi}), to the final expression
\be
\dfrac{\partial n^{\bxi}}{\partial
\xi_\lambda}=\left(\hat{1}+\chi^{\bxi}\star f^{\bxi}_{\rm Hxc}\right)\star 
\Delta
n^{\bxi}_{{\rm s},\lambda},
\ee
or, more explicitly,
\be\label{eq:final_expression_deriv_ens_dens_wrt_weight_lambda}
\begin{aligned}
\dfrac{\partial n^{\bxi}({\br})}{\partial
\xi_\lambda}&=\Delta
n^{\bxi}_{{\rm s},\lambda}({\br})
\\
&\quad+\int d{\br}'\int d{\br}" \chi^{\bxi}({\br},{\br}')f^{\bxi}_{\rm
Hxc}({\br}',{\br}")\Delta
n^{\bxi}_{{\rm s},\lambda}({\br}"),
\end{aligned}
\ee
which is our third and last exact key result.\\

In summary, applying the above equation (where the physical (static) ensemble density-density
linear response function $\chi^{\bxi}$ has been determined beforehand,
{\it via} the Dyson Eq.~(\ref{eq:dyson_n_equal_nxi})) in conjunction with
Eqs.~(\ref{eq:gathering_KS_terms_deriv_ens_dens_wrt_weight}) and
(\ref{eq:ind_dens_from_ens_dens_phys_main_text}) gives directly access to both
ground- and excited-state physical densities. 
\subsection{Summary and key conclusions of the exact theory}

An exact KS-like ensemble density functional has been derived for any 
(ground or excited) energy level in Eq.~(\ref{eq:ind_ener_ens_func}).
Writing its stationarity at the exact ensemble density is equivalent to
correcting the stationarity of the KS energy level (within the ensemble)
with additional terms that describe the discrepancy in individual
densities between the KS ensemble and the true physical one, as depicted
in Eqs.~(\ref{eq:stat_cond_true_from_KS}) and
(\ref{eq:true_dens_from_stat_cond}). The latter feature of the present
theory is equivalently
referred to as describing density-driven correlations in
ensembles~\cite{PRL19_Gould_DD_correlation,Fromager_2020,Cernatic2022,gould2024stationaryconditionsexcitedstates}. Finally, the working Eqs.~(\ref{eq:ind_dens_from_ens_dens_phys_main_text})
and (\ref{eq:final_expression_deriv_ens_dens_wrt_weight_lambda})
have been derived in order to evaluate, in principle exactly, from the KS ensemble and the true physical
ensemble density-density linear response function, any ground or excited
density.

\section{State-specific KS schemes relying on ensemble
density-functional approximations}\label{sec:ind_KS_pots_from_ens_DFAs}

The purpose of this section is to identify ensemble
density-functional approximations that, when applied to the exact
state-specific stationarity condition of
Eq.~(\ref{eq:final_ind_state_stat_cond}), lead to a KS-like equation for
the targeted state.  

\subsection{Exact Hartree-exchange-only approximation
}\label{sec:ind_exact_Hx_approximation}


The first approximation we can think of, which is likely to lead to a state-specific KS
equation in the present context, is the Hx-only one~\cite{Fromager_2020}. Often referred to
as ensemble exact exchange (EEXX), it simply consists in neglecting the
ensemble density-functional correlation energy ($E^{\bxi}_{\rm
Hxc}[n]\approx E^{\bxi}_{\rm
Hx}[n]$) in the exact theory. In particular, as density-driven correlations~\cite{PRL19_Gould_DD_correlation,Fromager_2020,Cernatic2022} are
ignored, there will be no need to distinguish the physical individual
densities from the KS ones, thus leading to substantial simplifications,
as shown in the following. We should stress that, even though the total
ensemble Hx functional has a clear (but somehow less trivial than in the
standard ground-state case~\cite{gould2017hartree}) definition, which can be written explicitly
as follows, in terms of the partially-interacting ensemble HK functional
(see Eq.~(\ref{eq:ens_part_int_decomp_ind_func})), 
\bse
\begin{align}
E^{\bxi}_{\rm
Hx}[n]&=\left.\dfrac{\partial F^{{\bxi},\alpha}[n]}{\partial
\alpha}\right|_{\alpha=0}
\\
\label{eq:exact_ens_Hx_func_exp}
&=\sum_{\nu\geq 0}\xi_\nu\innerop{\Phi^{\bxi}_{\nu}[n]}{\hat{W}_{\rm
ee}}{\Phi^{\bxi}_{\nu}[n]},
\end{align}
\ese
where, as expected from degenerate-perturbation theory through first
order (in the interaction strength $\alpha$), the KS wavefunctions are not necessarily single
determinants, the decomposition into Hartree and exchange contributions
is not unique~\cite{PRL20_Gould_Hartree_def_from_ACDF_th,Cernatic2022}.
For the sake of generality, we will keep both terms together and rely on
the expression of Eq.~(\ref{eq:exact_ens_Hx_func_exp}) for our
purpose.\\

If we now proceed in Eq.~(\ref{eq:final_ind_state_stat_cond}) with the
following simplifications (that hold at the Hx-only level of
approximation), 
\begin{subequations}\label{eq:collection_Hx-only_approximations}
\begin{align}
\label{eq:Hx_approx_wf}
\Psi^{{\bxi}}_\nu[n]&\approx \Phi^{{\bxi}}_\nu[n],
\\
v_{\rm s}^{{\bxi}}[n]&\approx v^{{\bxi}}[n]+v^{\bxi}_{\rm Hx}[n] 
\underset{n=n^{\bxi}}{=:} v_{\rm ext}+v^{\bxi}_{\rm Hx},
\\
f^{\bxi}_{\rm Hxc}[n]&\approx f^{\bxi}_{\rm
Hx}[n]\equiv \dfrac{\delta v^{\bxi}_{\rm Hx}[n]}{\delta n}\underset{n=n^{\bxi}}{=:}f^{\bxi}_{\rm 
Hx}, 
\end{align}
\end{subequations}
the stationarity condition for the targeted state $\nu$ reads
\begin{subequations}\label{eq:Hx_only_stat_cond_step1}
\begin{align}
0&\approx\Bigg[\delta T_{{\rm
s},\nu}^{\bxi}[n]
+\contract{v_{\rm ext}+v^{\bxi}_{\rm Hx}
}{\delta n_{\Phi^{{\bxi}}_\nu[n]}}
\\
&
\quad+
\sum_{\lambda>0}\left(\delta_{\lambda\nu}-\xi_\lambda\right)
\contract{\delta n}{\dfrac{\partial v^{\bxi}_{\rm Hx}[n]}{\partial
\xi_\lambda}}
\Bigg]_{n=n^{\bxi}}
\\
&
\quad
+\contract{\delta n}{f^{\bxi}_{\rm
Hx}
\star\left(n_{\Phi^{{\bxi}}_\nu}-n^{\bxi}\right)}
,
\end{align}
\end{subequations}
or, in a more compact form,
\be\label{eq:Hx_stat_cond_simp}
\begin{aligned}
\delta T_{{\rm
s},\nu}^{\bxi}[n]
&+\contract{\delta n_{\Phi^{{\bxi}}_\nu[n]}}{v_{\rm ext}+v^{\bxi}_{\rm Hx}
}
\\
&
+\contract{\delta n}{v^{\bxi}_{{\rm Hx},\nu}-v^{\bxi}_{\rm Hx}}
\underset{n=n^{\bxi}}{\approx}
0,
\end{aligned}
\ee
where the individual Hx potential $v^{\bxi}_{{\rm Hx},\nu}$ is defined
as follows, 
\be\label{eq:ind_Hx_pot_def}
v^{\bxi}_{{\rm Hx},\nu}:=v^{\bxi}_{\rm Hx}+\sum_{\lambda>0}\left(\delta_{\lambda\nu}-\xi_\lambda\right)
\left.\dfrac{\partial
}{\partial
\xi_\lambda}\left(v^{\bxi}_{\rm Hx}[n_{\rm s}^{\bxi,\bze}]\right)
\right|_{\bze=\bxi},
\ee
and the double-weight ensemble KS density reads (note that $n_{\rm s}^{\bxi,\bze=\bxi}=n_{\rm
s}^{\bxi}=n^{\bxi}$) 
\be
n_{\rm s}^{\bxi,\bze}=\sum_{\kappa\geq 0}\xi_\kappa
n_{\Phi^{{\bze}}_\kappa}
=n_{\Phi^{{\bze}}_0}+\sum_{\lambda>0}\xi_\lambda
\left(n_{\Phi^{{\bze}}_\lambda}-n_{\Phi^{{\bze}}_0}\right).
\ee
The equivalence between Eqs.~(\ref{eq:Hx_only_stat_cond_step1}) and
(\ref{eq:Hx_stat_cond_simp}) simply comes from the
following equality:

\be\label{eq:extraction_ind_Hx_pot}
\begin{aligned}
&\sum_{\lambda>0}\left(\delta_{\lambda\nu}-\xi_\lambda\right)
\left.\dfrac{\partial
}{\partial
\xi_\lambda}\left(v^{\bxi}_{\rm Hx}[n_{\rm s}^{\bxi,\bze}]\right)
\right|_{\bze=\bxi}
\\
&=\sum_{\lambda>0}\left(\delta_{\lambda\nu}-\xi_\lambda\right)
\left.\dfrac{\partial v^{\bxi}_{\rm Hx}[n]}{\partial
\xi_\lambda}\right|_{n=n^{\bxi}}
\\
&
\quad
+f^{\bxi}_{\rm
Hx}\star\left(n_{\Phi^{{\bxi}}_\nu}-n^{\bxi}\right).
\end{aligned}
\ee
Note that Eq.~(\ref{eq:ind_Hx_pot_def}) echoes the ``exact'' expression of
Eq.~(\ref{eq:def_ind_Hxc_pot}). The fact that the weight-dependence of
the KS wavefunctions is neglected implies the neglect of density-driven
correlations, as readily seen from
Eq.~(\ref{eq:ind_state_dens_extraction_PRL}).\\

Returning to the Hx-only stationarity condition of
Eq.~(\ref{eq:Hx_stat_cond_simp}), we still need to connect the ensemble
density variation $\delta n$ to that of the KS state of interest if we
want to recover a KS-like equation. For that purpose, let us 
differentiate the ensemble density constraint of
Eq.~({\ref{eq:ens_KS_dens_constraint}}) as follows,
\be
\dfrac{\partial n}{\partial \xi_\lambda}=0=
n_{\Phi^{{\bxi}}_\lambda[n]}-n_{\Phi^{{\bxi}}_0[n]}
+\sum_{\kappa\geq 0}\xi_\kappa \dfrac{\partial
n_{\Phi^{{\bxi}}_\kappa[n]}}{\partial \xi_\lambda},
\ee
which leads to the explicit relation:
\begin{subequations}
\begin{align}
n_{\Phi^{{\bxi}}_\nu[n]}&=n+\sum_{\lambda>0}\left(\delta_{\lambda\nu}-\xi_\lambda\right)\left(n_{\Phi^{{\bxi}}_\lambda[n]}-n_{\Phi^{{\bxi}}_0[n]}\right)
\\
\label{eq:general_ind_state_dens_from_ens_dens_n}
&=
n-\sum_{\lambda>0,\kappa\geq 0}\left(\delta_{\lambda\nu}-\xi_\lambda\right)\xi_\kappa \dfrac{\partial
n_{\Phi^{{\bxi}}_\kappa[n]}}{\partial \xi_\lambda}.
\end{align}
\end{subequations}
By neglecting the weight dependence of all individual KS density
variations within the ensemble, {\ie}, 
\be
\dfrac{\partial
}{\partial \xi_\lambda}
\left(
\delta n_{\Phi^{{\bxi}}_\kappa[n]}
\right)
\underset{\kappa\geq 0}{\approx} 0,
\ee
it immediately follows that
\be
\delta n_{\Phi^{{\bxi}}_\nu[n]}\approx \delta n,
\ee
and 
Eq.~(\ref{eq:Hx_stat_cond_simp}) becomes
\be
\delta T_{{\rm
s},\nu}^{\bxi}[n]
+\contract{\delta n_{\Phi^{{\bxi}}_\nu[n]}}{v_{\rm ext}+v^{\bxi}_{{\rm Hx},\nu}
}
\underset{n=n^{\bxi}}{\approx}0,
\ee
or, equivalently (see Eq.~(\ref{eq:ens_Ts_func_ind_state_def})),
\be\label{eq:final_form_stat_cond_Hx-only}
\innerop{\delta
\Phi^{{\bxi}}_\nu[n]}{\hat{T}+\hat{V}_{\rm ext}+
\hat{V}^{\bxi}_{{\rm Hx},\nu}}{\Phi^{{\bxi}}_\nu}
\underset{n=n^{\bxi}}{\approx} 0,
\ee
where $\hat{V}^{\bxi}_{{\rm Hx},\nu}=\int d\br\,v^{\bxi}_{{\rm
Hx},\nu}(\br)\,\hat{n}(\br)$.\\

In conclusion, it is {\it sufficient} to solve the following state-specific Hx-only KS equation,  
\be\label{eq:Hx-only_state_specific_KSE}
\left[\hat{T}+\hat{V}_{\rm ext}+
\hat{V}^{\bxi}_{{\rm Hx},\nu}
\right]\ket{\tilde{\Phi}^{{\bxi}}_\nu}
=\tilde{\mathcal{E}}^{\bxi}_\nu\ket{\tilde{\Phi}^{{\bxi}}_\nu},
\ee
in order to satisfy the (approximate) stationarity condition of
Eq.~(\ref{eq:final_form_stat_cond_Hx-only}). It becomes {\it necessary} if the full $N$-electron Hilbert space
can be spanned with $\ket{\delta \Phi^{{\bxi}}_\nu[n]}$ through
infinitesimal variations of the ensemble density $\delta n$ around the
exact one $n^{\bxi}$. Most importantly,
Eq.~(\ref{eq:Hx-only_state_specific_KSE}) suggests an alternative ensemble-based
but state-specific construction of KS wavefunctions for (ground and) excited states that is
expected to provide, at least in principle, a better description of each
state individually. The inclusion of ensemble correlation effects into
such a construction is {\it a priori} not trivial, as shown in
Sec.~\ref{sec:ind_state_KSDFT}, and is left for future
work.\\       

For the sake of completeness, let us finally evaluate the corresponding Hx-only energy level
expression. Combining the exact 
Eqs.~(\ref{eq:ind_nu_energy_dens_func_for_exact_ens_dens}) and (\ref{eq:ind_ener_ens_func}) with the Hx-only approximations of
Eq.~(\ref{eq:collection_Hx-only_approximations}) leads to 
\be
\begin{aligned}
&E_\nu
\approx \innerop{\Phi^{{\bxi}}_\nu}{\hat{T}+\hat{V}_{\rm
ext}}{\Phi^{{\bxi}}_\nu}+E^{\bxi}_{\rm Hx}[n^{\bxi}]
\\
&+\sum_{\lambda>0}\left(\delta_{\lambda\nu}-\xi_\lambda\right)\left.\dfrac{\partial E^{\bxi}_{\rm Hx}[n]}{\partial
\xi_\lambda}\right|_{n=n^{\bxi}}+\contract{v^{\bxi}_{\rm
Hx}}{n_{\Phi^{{\bxi}}_\nu}-n^{\bxi}},
\end{aligned}
\ee 
or, equivalently, by analogy with Eq.~(\ref{eq:extraction_ind_Hx_pot}),
\be
\begin{aligned}
E_\nu
&\approx \innerop{\Phi^{{\bxi}}_\nu}{\hat{T}+\hat{V}_{\rm
ext}}{\Phi^{{\bxi}}_\nu}+E^{\bxi}_{\rm Hx}[n^{\bxi}]
\\
&\quad+\sum_{\lambda>0}\left(\delta_{\lambda\nu}-\xi_\lambda\right)\left.\dfrac{\partial }{\partial
\xi_\lambda}\left(E^{\bxi}_{\rm Hx}[n_{\rm s}^{\bxi,\bze}]\right)\right|_{\bze=\bxi}
.
\end{aligned}
\ee 
Applying the following trick~\cite{Fromager_2020}, 
\be
\begin{aligned}
&\left.\dfrac{\partial }{\partial
\xi_\lambda}\left(E^{\bxi}_{\rm Hx}[n_{\rm s}^{\bxi,\bze}]\right)\right|_{\bze=\bxi}
\\
&=\dfrac{\partial }{\partial
\xi_\lambda}\left(E^{\bxi}_{\rm Hx}[n_{\rm s}^{\bxi,\bxi}]\right)
-\left.\dfrac{\partial }{\partial
\xi_\lambda}\left(E^{\bze}_{\rm Hx}[n_{\rm
s}^{\bze,\bxi}]\right)\right|_{\bze=\bxi},
\end{aligned}
\ee
where, according to Eq.~(\ref{eq:exact_ens_Hx_func_exp}),
\be
E^{\bze}_{\rm Hx}[n_{\rm
s}^{\bze,\bxi}]=\sum_{\kappa\geq0}\zeta_\kappa\innerop{\Phi_\kappa^{\bxi}}{\hat{W}_{\rm
ee}}{\Phi_\kappa^{\bxi}},
\ee
finally gives back the expected Hx-only energy level
expression~\cite{Fromager_2020}, 
\be
E_\nu\approx \innerop{\Phi^{{\bxi}}_\nu}{\hat{T}+\hat{W}_{\rm ee}+\hat{V}_{\rm
ext}
}{\Phi^{{\bxi}}_\nu}.
\ee
Replacing in the above equation $\Phi^{{\bxi}}_\nu$ by the state-specific KS
wavefunction $\tilde{\Phi}^{{\bxi}}_\nu$ defined in
Eq.~(\ref{eq:Hx-only_state_specific_KSE}) may provide a
better starting point in the description of the $\nu$th energy level.
The implementation and calibration of such a scheme from the EEXX
functional is left for future work. 

\subsection{Connecting ensemble DFT to orbital-optimized DFT of excited
states}\label{sec:OO-DFT_excited-states}

While the computation of excited-state energies from the stationary
points 
to the regular (ground-state) KS energy functional has gained an
increasing attention~\cite{Levi20_Variational,Ivanov21_Method,Hait21_Orbital,Schmerwitz22_Variational}, its formal connection to ensemble DFT has
only been invoked recently in the
context of ensemble potential functional
theory~\cite{gould2024stationaryconditionsexcitedstates}. Such a
connection can also be established within the present density-functional
formalism. Indeed, from the exact ensemble density-functional energy level
expression of Eq.~(\ref{eq:ind_ener_ens_func}) and the following
ensemble density-functional approximations,             
\begin{subequations}
\begin{align}
\label{eq:drastic_approx_all_densities_equal}
n_{\Psi^{{\bxi}}_\nu[n]}(\br)&\approx n(\br)\approx
n_{\Phi^{{\bxi}}_\nu[n]}(\br),
\\
E^{\bxi}_{\rm Hxc}[n]&\approx \sum_{\nu\geq 0}\xi_\nu\,E_{\rm
Hxc}\left[n_{\Phi^{{\bxi}}_\nu[n]}\right],
\end{align}
\end{subequations}
where $E_{\rm Hxc}[n]=E^{{\bxi}=0}_{\rm Hxc}[n]$ is the standard
ground-state Hxc density functional, we obtain a drastically
simplified energy expression 
\bse\label{eq:ind_ener_ens_func_approx2}
\begin{align}
E^{\bxi}_{\nu}[n]&\approx
\innerop{\Phi^{{\bxi}}_\nu[n]}{\hat{T}+\hat{V}_{\rm ext}}{\Phi^{{\bxi}}_\nu[n]} +E_{\rm
Hxc}\left[n_{\Phi^{{\bxi}}_\nu[n]}\right]
\\
&=E^{\rm KS-DFT}\left[\Phi^{{\bxi}}_\nu[n]\right]
,
\end{align}
\ese
where the ensemble-density functional KS wavefunction
$\Phi^{{\bxi}}_\nu[n]$ is simply inserted into the regular KS-DFT
energy functional   
\be
E^{\rm KS-DFT}\left[\Phi\right]=\innerop{\Phi}{\hat{T}+\hat{V}_{\rm ext}
}{\Phi}
+E_{\rm
Hxc}[n_\Phi].
\ee
Note the drastic approximation that is made in
Eq.~(\ref{eq:drastic_approx_all_densities_equal}), which implies in
particular a total neglect of density-driven correlations. It also means that
$\partial n_{\Phi^{{\bxi}}_\nu[n]}/\partial \xi_\lambda\approx 0$, thus
leading to Eq.~(\ref{eq:ind_ener_ens_func_approx2}). Consequently, the
stationarity of the ensemble density-functional energy level now reads  
\be
0\approx \left.\innerop{\delta
\Phi^{{\bxi}}_\nu[n]}{\hat{T}+\hat{V}_{\rm ext}+
\hat{V}_{{\rm Hxc}}\left[n_{\Phi^{{\bxi}}_\nu}\right]}{\Phi^{{\bxi}}_\nu}
\right|_{n=n^{\bxi}},
\ee
where $\hat{V}_{{\rm Hxc}}[n]=\int d\br\,\delta E_{\rm Hxc}[n]/\delta
n({\br})\,\hat{n}(\br)$ is the regular ground-state Hxc potential
operator. The above condition is fulfilled by any ground
or excited self-consistent solution to the KS equation, 
\be
\left[\hat{T}+\hat{V}_{\rm ext}+
\hat{V}_{{\rm
Hxc}}\left[n_{{\Phi}_\nu}\right]\right]\myket{\Phi_\nu}=\mathcal{E}_\nu\myket{\Phi_\nu},
\ee
which is equivalent to
\be
\Phi_\nu=\underset{\Phi}{\arg {\rm
stat}}
\left\{E^{\rm KS-DFT}\left[\Phi\right]\right\}.
\ee
Using $\Phi_\nu$ (instead of the ensemble-based KS
wavefunction $\Phi^{\bxi}_\nu$) as reference in the exact derivation of 
individual KS-like ensemble density-functional stationarity conditions (see
Eq.~(\ref{eq:final_ind_state_stat_cond})) would be a
natural follow-up. It would provide a formal exactification (from the
perspective of ensemble DFT) of
orbital-optimized DFT for excited states and, most importantly, it may
serve as a guide for converging towards the desired state and
possibly also for improving its energy level in terms of accuracy. Work is
currently in progress in this direction. 

\section{Conclusions and perspectives}\label{sec:conclusions}

An in-principle exact ensemble DFT of energy levels has been derived.
Our two first key results are the expression, in terms of the
(weight-dependent) Hxc ensemble density
functional, of an exact state-specific KS energy expression (in
Eq.~(\ref{eq:ind_ener_ens_func})) and the
resulting stationarity condition (in
Eq.~(\ref{eq:stat_cond_true_from_KS})). We have shown that, when the ensemble
density is mapped onto the KS ensemble, which does not guarantee that interacting and noninteracting densities match individually, the exact
individual densities can still be recovered from the theory, through the
stationarity of the targeted ensemble density-functional energy level. 
In addition, and this is our third exact key result, working equations
have been derived (in Eq.~(\ref{eq:final_expression_deriv_ens_dens_wrt_weight_lambda})) for evaluating, in
principle exactly, the true physical densities from the (static)
ensemble density-density linear response function. Starting from the
exact theory, two approximate schemes have been formulated on the basis
of well-identified ensemble density-functional approximations. The first
one relies on the exact ensemble Hx functional from which
a state-specific Hx potential naturally emerges. In the second scheme, the
ensemble Hxc functional is (approximately) constructed from the regular
(ground-state) Hxc one and the individual KS densities. If, in addition, we assume that all
densities within the ensemble are similar (which is a crude
approximation), the orbital-optimized DFT of excited states~\cite{Levi20_Variational,Ivanov21_Method,Hait21_Orbital,Schmerwitz22_Variational} is
immediately
recovered. Either scheme could actually be used as starting point in the exact
theory, instead of the
conventional ensemble KS one, in order to better describe each state
individually. In practice, one could also benefit from recent advances
in the development of ensemble density functionals~\cite{gould2024excitationenergiesstatespecificensemble} in order to
model the missing Hxc energy contributions. In conclusion, the present work opens up new theoretical and
computational perspectives in the field of DFT for excited states.

\section*{Acknowledgements}

The author would like to thank Hardy Gross for the inspiring talk he
gave in July 2024 (at the first ensemble DFT workshop in Durham, 
UK) on the exact evaluation of Fukui functions in DFT. The author is also very happy to contribute
with the present paper to a special issue published in honor of Trygve Helgaker.
Trygve has always been a great source of inspiration. The author is very
grateful to him, in particular for his mathematical work on
DFT.  
\appendix

\section{Connecting individual-state and ensemble
functionals}\label{sec:ind_from_ens_func}

Let us introduce, for the sake of generality, the following partially-interacting
version of the ensemble HK functional,
\be\label{eq:ens_part_int_decomp_ind_func}
\begin{aligned}
F^{{\bxi},\alpha}[n]&=\sum_{\nu\geq
0}\xi_\nu\,F^{{\bxi},\alpha}_{\nu}[n]
\\
&=F^{{\bxi},\alpha}_{0}[n]+\sum_{\lambda>0}\xi_\lambda\left(F^{{\bxi},\alpha}_{\lambda}[n]-F^{{\bxi},\alpha}_{0}[n]\right)
,
\end{aligned}
\ee
where
\be\label{eq:ind_partinteracting_HK_func}
F^{{\bxi},\alpha}_{\nu}[n]=\mybra{\Psi^{{\bxi},\alpha}_\nu[n]}\hat{T}+\alpha\hat{W}_{\rm ee}\myket{\Psi^{{\bxi},\alpha}_\nu[n]}
\ee
and $\Psi^{{\bxi},\alpha}_\nu[n]$ is the $\nu$th solution to the
following partially-interacting ($0\leq \alpha\leq 1$) ensemble
density-functional Schr\"{o}dinger equation: 
\be\label{eq:part_interacting_ens_dens_SE}
\hat{H}^{{\bxi},\alpha}[n]
\myket{\Psi^{{\bxi},\alpha}_\nu[n]}
{=}\check{E}^{{\bxi},\alpha}_\nu[n]\myket{\Psi^{{\bxi},\alpha}_\nu[n]},
\ee
with 
\be\label{eq:partint_ens_dens_func_Hamil}
\hat{H}^{{\bxi},\alpha}[n]=\hat{T}+\alpha\hat{W}_{\rm ee}+\int d\br\,
v^{{\bxi},\alpha}[n](\br)\,\hat{n}(\br).
\ee
The local potential $v^{{\bxi},\alpha}[n]$ is adjusted such that the
following ensemble density constraint is fulfilled for any value of
$\alpha$ in the range $0\leq \alpha\leq 1$:
\be\label{eq:partint_ens_dens_constraint}
\begin{aligned}
n&=\sum_{\nu\geq 0}\xi_\nu\,n_{\Psi^{{\bxi},\alpha}_\nu[n]}
\\
&=n_{\Psi^{{\bxi},\alpha}_0[n]}+\sum_{\lambda>0}\xi_\lambda\left(n_{\Psi^{{\bxi},\alpha}_\lambda[n]}-n_{\Psi^{{\bxi},\alpha}_0[n]}\right)
.
\end{aligned}
\ee

If we denote 
\be\label{eq:pot_func_ens_ener_decomp}
\begin{aligned}
E^{{{\bxi},\alpha}}[v]&=\sum_{\nu\geq 0}\xi_\nu\,E^{\alpha}_{\nu}[v]
\\
&=E^{\alpha}_{0}[v]+\sum_{\lambda>0}\xi_\lambda\left(E^{\alpha}_{\lambda}[v]-E^{\alpha}_{0}[v]\right)
\end{aligned}
\ee
the exact ensemble energy of the partially-interacting Hamiltonian $\hat{T}+\alpha\hat{W}_{\rm ee}+\int d\br\,v(\br)\hat{n}(\br)$ for a
given local potential $v$, $E^{\alpha}_{\nu}[v]$ being the $\nu$th
eigenvalue of the latter Hamiltonian, it comes
\begin{subequations}
\begin{align}
E^{{{\bxi},\alpha}}[v]&=\min_{n}\left\{F^{{\bxi},\alpha}[n]+\contract{v}{n}\right\},\forall v
\\
&\leq F^{{\bxi},\alpha}[n]+\contract{v}{n},\forall n,\forall v,
\end{align}
\end{subequations}
or, equivalently,
\be
F^{{\bxi},\alpha}[n]\geq E^{{{\bxi},\alpha}}[v]-\contract{v}{n},\forall n,\forall v,
\ee
thus leading to the following Levy--Lieb
expression~\cite{LFTransform-Lieb,Borgoo15_Excitation,Cernatic2022}
\be\label{eq:Lieb_max_part_int_ens}
F^{{\bxi},\alpha}[n]=\max_{v}\left\{E^{{{\bxi},\alpha}}[v]-\contract{v}{n}\right\},
\forall n.
\ee
Note that, according to Eqs.~(\ref{eq:ens_part_int_decomp_ind_func}) and (\ref{eq:ind_partinteracting_HK_func}), the maximizing potential in
Eq.~(\ref{eq:Lieb_max_part_int_ens}) is $v^{{\bxi},\alpha}[n]$.
Therefore, 
\be\label{eq:max_partint_pot_func_deriv_exp}
v^{{\bxi},\alpha}[n]=-\dfrac{\delta F^{{\bxi},\alpha}[n]}{\delta
n}.
\ee

Let us now focus on the extraction of the individual-state functional $F^{{\bxi},\alpha}_{\nu}[n]$ from
the ensemble functional $F^{{\bxi},\alpha}[n]$. By analogy with the
extraction of exact energy levels from the KS
ensemble~\cite{deur2019ground,Fromager_2020}, we will use first order
derivatives in the ensemble weights for that purpose. 
According to Eqs.~(\ref{eq:pot_func_ens_ener_decomp}) and
(\ref{eq:Lieb_max_part_int_ens}), 
\begin{subequations}
\begin{align}
\dfrac{\partial F^{{\bxi},\alpha}[n]}{\partial
\xi_\lambda}&\underset{\lambda>0}{=}\left.\dfrac{\partial
E^{{{\bxi},\alpha}}[v]}{\partial \xi_\lambda}\right|_{v=v^{{\bxi},\alpha}[n]}
\\
&=\left.\left(E^{\alpha}_{\lambda}[v]-E^{\alpha}_{0}[v]\right)\right|_{v=v^{{\bxi},\alpha}[n]},
\end{align}
\end{subequations}
thus leading to (see Eqs.~(\ref{eq:part_interacting_ens_dens_SE}) and (\ref{eq:partint_ens_dens_func_Hamil}))
\be
\dfrac{\partial F^{{\bxi},\alpha}[n]}{\partial
\xi_\lambda}&\underset{\lambda>0}{=}\check{E}^{{\bxi},\alpha}_\lambda[n]-\check{E}^{{\bxi},\alpha}_0[n],
\ee
or, equivalently (see Eq.~(\ref{eq:ind_partinteracting_HK_func})),
\be\label{eq:dFens_part_int_over_dxinu}
\begin{aligned}
\dfrac{\partial F^{{\bxi},\alpha}[n]}{\partial
\xi_\lambda}&=F^{{\bxi},\alpha}_{\lambda}[n]-F^{{\bxi},\alpha}_{0}[n]
\\
&\quad+\contract{v^{{\bxi},\alpha}[n]}{n_{\Psi^{{\bxi},\alpha}_\lambda[n]}-n_{\Psi^{{\bxi},\alpha}_0[n]}}.
\end{aligned}
\ee
Since, for any (ground or excited) state $\nu$,
\be
\begin{aligned}
F^{{\bxi},\alpha}_{\nu}[n]&\underset{\nu\geq
0}{=}
F^{{\bxi},\alpha}_{0}[n]
\\
&\quad\quad+\sum_{\lambda>0}\delta_{\lambda\nu}\left(F^{{\bxi},\alpha}_{\lambda}[n]-F^{{\bxi},\alpha}_{0}[n]\right),
\end{aligned}
\ee
thus leading to, according to Eq.~(\ref{eq:ens_part_int_decomp_ind_func}),
\be
\begin{aligned}
&F^{{\bxi},\alpha}_{\nu}[n]\underset{\nu\geq
0}{=}F^{{\bxi},\alpha}[n]
\\
&+\sum_{\lambda>0}\left(\delta_{\lambda\nu}-\xi_\lambda\right)\times\left(F^{{\bxi},\alpha}_{\lambda}[n]-F^{{\bxi},\alpha}_{0}[n]\right),
\end{aligned}
\ee
it comes from Eq.~(\ref{eq:dFens_part_int_over_dxinu}) and the ensemble density
constraint of Eq.~(\ref{eq:partint_ens_dens_constraint}),
\be
\begin{aligned}
F^{{\bxi},\alpha}_{\nu}[n]
&\underset{\nu\geq
0}{=}
F^{{\bxi},\alpha}[n]
+\sum_{\lambda>0}\left(\delta_{\lambda\nu}-\xi_\lambda\right)\dfrac{\partial F^{{\bxi},\alpha}[n]}{\partial
\xi_\lambda}
\\
&\quad\quad-\contract{v^{{\bxi},\alpha}[n]}
{n_{\Psi^{{\bxi},\alpha}_\nu[n]}-n_{\Psi^{{\bxi},\alpha}_0[n]}}
\\
&\quad\quad+\contract{v^{{\bxi},\alpha}[n]}
{n-n_{\Psi^{{\bxi},\alpha}_0[n]}},
\end{aligned}
\ee
which gives, according to Eq.~(\ref{eq:max_partint_pot_func_deriv_exp}), the
final expression 
\be
\begin{aligned}
F^{{\bxi},\alpha}_{\nu}[n]
&\underset{\nu\geq
0}{=}
F^{{\bxi},\alpha}[n]
+\sum_{\lambda>0}\left(\delta_{\lambda\nu}-\xi_\lambda\right)\dfrac{\partial F^{{\bxi},\alpha}[n]}{\partial
\xi_\lambda}
\\
&\quad\quad-\contract{\dfrac{\delta F^{{\bxi},\alpha}[n]}{\delta
n}}{n-n_{\Psi^{{\bxi},\alpha}_\nu[n]}}.
\end{aligned}
\ee
Consequently, the true interacting ($\alpha=1$) and noninteracting
kinetic energy ($\alpha=0$) individual functionals can be connected to
the ensemble ones as follows (see Eqs.~(\ref{eq:interacting_ens_dens_SE}) and (\ref{eq:ens_dens_KSE})), 
\be
\begin{aligned}
F^{{\bxi}}_{\nu}[n]&=F^{{\bxi},\alpha=1}_{\nu}[n]
\\
&
\underset{\nu\geq
0}{=}
F^{{\bxi}}[n]
+\sum_{\lambda>0}\left(\delta_{\lambda\nu}-\xi_\lambda\right)\dfrac{\partial F^{{\bxi}}[n]}{\partial
\xi_\lambda}
\\
&\quad\quad-\contract{\dfrac{\delta F^{{\bxi}}[n]}{\delta
n}}{n-n_{\Psi^{{\bxi}}_\nu[n]}},
\end{aligned}
\ee
and
\be
\begin{aligned}
T_{{\rm s},\nu}^{\bxi}[n] 
\underset{\nu\geq
0}{=}&
T_{{\rm s}}^{\bxi}[n]
+\sum_{\lambda>0}\left(\delta_{\lambda\nu}-\xi_\lambda\right)\dfrac{\partial
T_{{\rm s}}^{\bxi}[n]}{\partial
\xi_\lambda}
\\
&-\contract{\dfrac{\delta T_{{\rm s}}^{\bxi}[n]}{\delta
n}}{n-n_{\Phi^{{\bxi}}_\nu[n]}},
\end{aligned}
\ee
respectively, thus leading, by difference (see
Eqs.~(\ref{eq:ens_KS_decomp}) and (\ref{eq:KS_decomp_Hxc_energy_level})), to the expression in
Eq.~(\ref{eq:ind_state_Hxc_func_from_ens_func}) of the individual Hxc
energy functional. 

\section{Simplification of the stationarity condition based on a single ensemble density-functional energy level
}\label{sec:appendix_simpl_stat_cond}

Starting from the stationarity condition of
Eq.~(\ref{eq:final_ind_state_stat_cond}), we first proceed with the
following simplification,
\begin{subequations}
\begin{align}
&\sum_{\lambda>0}\left(\delta_{\lambda\nu}-\xi_\lambda\right)\left.\dfrac{\partial
v^{\bxi}_{\rm Hxc}[n]}{\partial
\xi_\lambda}\right|_{n=n^{\bxi}}
\\
&=\sum_{\lambda>0}\left(\delta_{\lambda\nu}-\xi_\lambda\right)
\\
&\quad\times
\left(\dfrac{\partial
}{\partial \xi_\lambda}
\left(v^{\bxi}_{\rm Hxc}[n^{\bxi}]\right)
-\left.\dfrac{\delta v^{\bxi}_{\rm Hxc}[n]}{\delta n}\right|_{n=n^{\bxi}}\star \dfrac{\partial
n^{\bxi}}{\partial \xi_\lambda}\right)
\\
&=\sum_{\lambda>0}\left(\delta_{\lambda\nu}-\xi_\lambda\right)\dfrac{\partial
v^{\bxi}_{\rm Hxc}}{\partial \xi_\lambda}
\\
&\quad-f^{\bxi}_{\rm Hxc}\star\sum_{\lambda>0}\left(\delta_{\lambda\nu}-\xi_\lambda\right)
 \dfrac{\partial
n^{\bxi}}{\partial \xi_\lambda}
\\
\label{eq:ind_stat_cond_final_exp_kernel_ind_dens_minus_ens_dens}
&=\sum_{\lambda>0}\left(\delta_{\lambda\nu}-\xi_\lambda\right)\dfrac{\partial
v^{\bxi}_{\rm Hxc}}{\partial \xi_\lambda}
-f^{\bxi}_{\rm Hxc}\star \left(n_{\Psi_\nu}-n^{\bxi}\right),
\end{align}
\end{subequations}
where we used the shorthand notation $v^{\bxi}_{\rm Hxc}:=v^{\bxi}_{\rm
Hxc}[n^{\bxi}]$ and, in Eq.~(\ref{eq:ind_stat_cond_final_exp_kernel_ind_dens_minus_ens_dens}), the fact that the true {\it
physical}
ensemble density is an affine function of the ensemble weights (see
Eq.~(\ref{eq:phys_ens_dens_exp_EX_weights}) and
Ref.~\citenum{Fromager_2020}):
\be\label{eq:ind_dens_extraction_linearity_appendix}
\begin{aligned}
n_{\Psi_\nu}&\underset{\nu\geq
0}{=}n_{\Psi_0}+\sum_{\lambda>0}\delta_{\lambda\nu}\left(n_{\Psi_\lambda}-n_{\Psi_0}\right)
\\
&=\left(n^{\bxi}-\sum_{\lambda>0}\xi_\lambda\left(n_{\Psi_\lambda}-n_{\Psi_0}\right)\right)
\\
&\quad+\sum_{\lambda>0}\delta_{\lambda\nu}\left(n_{\Psi_\lambda}-n_{\Psi_0}\right)
\\
&=n^{\bxi}+\sum_{\lambda>0}\left(\delta_{\lambda\nu}-\xi_\lambda\right)\dfrac{\partial
n^{\bxi}}{\partial \xi_\lambda}.
\end{aligned}
\ee 
Moreover, if we now introduce the interacting
$\chi^{\bxi}[n]={\delta n}/{\delta v^{\bxi}[n]}: (\br,\br')\mapsto
{\delta n (\br)}/{\delta v^{\bxi}[n]}(\br')$ and non-interacting KS
$\chi^{\bxi}_{\rm s}[n]={\delta n}/{\delta v_{\rm s}^{\bxi}[n]}$
ensemble density functional density-density linear response functions,
it comes  
\be
\begin{aligned}
&-\dfrac{\delta v^{{\bxi}}[n]}{\delta
n}\star n_{\Psi_\nu}
+\dfrac{\delta v_{\rm s}^{{\bxi}}[n]}{\delta
n}\star n_{\Phi^{{\bxi}}_\nu}
\\
&=-\left[\chi^{\bxi}[n]\right]^{-1}\star n_{\Psi_\nu}+\left[\chi_{\rm
s}^{\bxi}[n]\right]^{-1}\star n_{\Phi^{{\bxi}}_\nu},
\end{aligned}
\ee
thus leading to, according to the Dyson Eq.~(\ref{eq:dyson_n_equal_nxi}),
\be\label{eq:simpl_pots_contributions_stat_cond_appendix}
\begin{aligned}
&\left[-\dfrac{\delta v^{{\bxi}}[n]}{\delta
n}\star n_{\Psi_\nu}
+\dfrac{\delta v_{\rm s}^{{\bxi}}[n]}{\delta
n}\star n_{\Phi^{{\bxi}}_\nu}\right]_{n=n^{\bxi}}
\\
&=\left[\chi_{\rm
s}^{\bxi}\right]^{-1}\star \left(n_{\Phi^{{\bxi}}_\nu}-n_{\Psi_\nu}\right)
+f^{\bxi}_{\rm Hxc}\star n_{\Psi_\nu}.
\end{aligned}
\ee
Combining
Eqs.~(\ref{eq:ind_stat_cond_final_exp_kernel_ind_dens_minus_ens_dens})
and (\ref{eq:simpl_pots_contributions_stat_cond_appendix}) finally gives
\be
\begin{aligned}
&\sum_{\lambda>0}\left(\delta_{\lambda\nu}-\xi_\lambda\right)\left.\dfrac{\partial
v^{\bxi}_{\rm Hxc}[n]}{\partial
\xi_\lambda}\right|_{n=n^{\bxi}}-f^{\bxi}_{\rm Hxc}\star n^{\bxi}
\\
&+\left[-\dfrac{\delta v^{{\bxi}}[n]}{\delta
n}\star n_{\Psi_\nu}
+\dfrac{\delta v_{\rm s}^{{\bxi}}[n]}{\delta
n}\star n_{\Phi^{{\bxi}}_\nu}\right]_{n=n^{\bxi}}
\\
&=
\sum_{\lambda>0}\left(\delta_{\lambda\nu}-\xi_\lambda\right)\dfrac{\partial
v^{\bxi}_{\rm Hxc}}{\partial \xi_\lambda}
+\left[\chi_{\rm
s}^{\bxi}\right]^{-1}\star
\left(n_{\Phi^{{\bxi}}_\nu}-n_{\Psi_\nu}\right),
\end{aligned}
\ee
which, when inserted into Eq.~(\ref{eq:final_ind_state_stat_cond}),
leads to Eq.~(\ref{eq:stat_cond_true_from_KS}).

\clearpage



\newcommand{\Aa}[0]{Aa}

\end{document}